\documentclass[12pt]{article}%
\usepackage{amsfonts}
\usepackage{multicol,subcaption}
\usepackage{afterpage}
\usepackage{amsmath,bbm,dsfont,mathrsfs,mathtools,appendix}
\usepackage{amssymb}
\usepackage{graphicx}
\usepackage{fullpage}%
\usepackage{framed}
\usepackage{xcolor}
\usepackage[colorlinks=true,linkcolor=blue,citecolor=green,plainpages=false,pdfpagelabels]%
{hyperref}%
\hypersetup{
	colorlinks,
	linkcolor={blue!60!green},
	citecolor={green!50!yellow!75!black},
	urlcolor={blue!80!black},
	linktoc=all
}
\providecommand{\U}[1]{\protect\rule{.1in}{.1in}}

\newtheorem{theorem}{Theorem}

\newtheorem{example}[theorem]{Example}

\newtheorem{lemma}[theorem]{Lemma}

\newtheorem{proposition}[theorem]{Proposition}
	
\newtheorem{remark}[theorem]{Remark}
\newtheorem{rem}{Remark}

\newenvironment{proof}[1][Proof]{\noindent\textbf{#1.} }{\ \rule{0.5em}{0.5em}}

\newcommand{\cF}{{\mathcal{F}}}
\newcommand{\infy}{\infty}
\newcommand{\id}{{\rm{id}}} %identity
\newcommand{\spec}{{\rm{spec}}}
\newcommand{\Spec}{{\rm{spec}}}
\newcommand{\res}{{\rm{res}}}
\newcommand{\cB}{{\mathcal{B}}}

\newcommand{\cH}{\mathcal{H}}
\newcommand{\R}{\mathbbm{R}}
\newcommand{\C}{\mathbb{C}}
\newcommand{\CC}{{\mathbb C}}
\newcommand{\N}{\mathbb{N}}
\newcommand{\cL}{\mathcal{L}}
\newcommand{\nn}{\nonumber}

\newcommand{\cT}{\mathcal{T}}

\newcommand{\F}{\mathcal{F}}

\newcommand{\cN}{\mathcal{N}}

\newcommand{\1}{\mathbbm{1}}
\newcommand{\cD}{\mathcal{D}}

\def\>{{\rangle}}
\def\<{{\langle}}
\newcommand{\be}{\begin{equation}}
	\newcommand{\ee}{\end{equation}}
\newcommand{\bea}{\begin{eqnarray}}
	\newcommand{\eea}{\end{eqnarray}}

\newcommand{\eps}{\varepsilon}

\newcommand{\floor}[1]{\left\lfloor {#1} \right\rfloor}

\newcommand{\ket}[1]{|#1\rangle} %ket
\newcommand{\bra}[1]{\langle#1|} %bra
\newcommand{\kb}[1]{|#1\rangle\!\langle#1|} %ketbra
\def\placeholder{\,\cdot\,}
\newcommand{\Tr}{\mathrm{Tr}}

\newcommand{\comment}[1]{}
\newcommand{\red}[1]{{\color{red} #1}}

\newcommand{\ran}{\operatorname{ran}}
\newcommand{\supp}{\operatorname{supp}}

\def\cK{\mathcal{K}}

\newcommand{\tr}{{\rm Tr}}

\numberwithin{equation}{section}
\numberwithin{theorem}{section}

\usepackage{color}
\definecolor{colorthree}{rgb}{0.01,0.51,0.93}

\def\smallsection#1{\bigskip\noindent\textbf{#1}}

\usepackage{setspace}

\newcommand{\footremember}[2]{%
    \footnote{#2}
    \newcounter{#1}
    \setcounter{#1}{\value{footnote}}%
}

\allowdisplaybreaks
\usepackage{subfiles}

\title{Quantitative Quantum Zeno and Strong Damping Limits in Strong Topology}

\author{Robert Salzmann \footremember{Cam}{Department of Applied Mathematics and Theoretical Physics, University of Cambridge, United Kingdom}\footremember{Lyon}{Univ Lyon, Inria, ENS Lyon, UCBL, LIP, F-69342, Lyon Cedex 07, France}}
\date{}

\begin{document}
%\frontmatter

\maketitle

\begin{abstract}
Frequent applications of a mixing quantum operation to a quantum system slow
down its time evolution and eventually drive it into the invariant subspace of the
    named operation. We prove this phenomenon, the quantum Zeno effect, and its continuous variant, strong damping, in a unified way for infinite-dimensional open quantum systems, while merely demanding that the respective mixing convergence holds pointwise for all states. Both results are quantitative in the following sense: Given the speed of convergence for the mixing limits, we can derive bounds on the convergence speed for the corresponding quantum Zeno and strong damping limits. We apply our results to prove quantum Zeno and strong damping limits for the photon loss channel with an explicit bound on the convergence speed.
\end{abstract}

\tableofcontents
%\mainmatter

\section{Introduction}

The \emph{quantum Zeno effect}, in its simplest form, describes the phenomenon in which frequent measurements of a quantum system can slow down its time evolution, eventually causing it to stop changing completely.
It has been extensively studied for closed quantum systems \cite{BN67,MisSud77,C72,F76,FP,ExIch05,ExIch21}, and for open quantum systems \cite{MaShvi03,BZ,BuFaNaPaYu18,MobWolf19,Ma04,G,BeckerDattaSalz_Zeno_2021,möbus2022optimal,Tim_TrotterZeno_2024} and experimental verification of the phenomenon was achieved in \cite{ItHeBoWi90,FiGuRai01,Streed_ContExperZeno_2006,schafer_experimentalZenoDym_2014}. 
The quantum Zeno effect has various applications for example in control of decoherence \cite{FJP04,HRBPK06}, quantum error correction \cite{EARV04,PSRDL12} and state preparation \cite{NTY03,NUY04,WYN08}.

Here, we consider the following general setup of the quantum Zeno effect in a possibly infinite-dimensional open quantum system evolving under a quantum dynamical semigroup, denoted by $\left(e^{t\cL}\right)_{t\ge 0},$ for some fixed time $t>0$ (cf. Section~\ref{sec:semigroups} for definition and discussion). During this time evolution a quantum operation\footnote{As defined in the Preliminaries, a quantum operation is a completely positive, trace non-increasing linear map.}, $M$, is applied $n$ times in 
equal intervals of time $t/n$.
The overall process is then given by the \emph{quantum Zeno product}
\begin{align}
\label{eq:IntroZenoProd}
    \left(Me^{t\cL/n}\right)^n.
\end{align}
The quantum operation $M$ is considered to be \emph{mixing}, in the sense that if applied multiple times it converges to its fixed point space. More precisely, for $P$ being the projection onto the fixed point space of $M$, we assume
    \begin{align}
    \label{eq:MixingOperationIntro}
        M^n\xrightarrow[n\to\infty]{} P,
    \end{align}
    in some reasonable topology.
We are then interested in the asymptotic behaviour of the quantum Zeno product \eqref{eq:IntroZenoProd} as the application frequency, $n$, of the quantum operation goes to infinity. The corresponding limit process, if it exists, is called the \emph{quantum Zeno limit.} \footnote{Henceforth, we often suppress the word ’quantum’ for simplicity.}

 We expect the Zeno product \eqref{eq:IntroZenoProd} to converge to an \emph{effective Zeno dynamics} on the fixed point space, i.e.
    \begin{align}
    \label{eq:ZenoConvIntro}
      \left(Me^{t\cL/n}\right)^n \xrightarrow[n\to\infty]{} e^{tP\cL P}P.
    \end{align}

    In this setup with a general semigroup dynamics with bounded generator $\cL$ and for quantum operations $M$ satisfying a certain spectral condition,\footnote{For a definition and discussion of named spectral condition please consider \cite{BeckerDattaSalz_Zeno_2021} or \cite[Section~4.1]{Salzmann_PhDthesis_2023}.} the quantum Zeno limit \eqref{eq:ZenoConvIntro} was proven to hold in \cite{MaShvi03,BZ,BuFaNaPaYu18,MobWolf19} and made quantitative in \cite{BeckerDattaSalz_Zeno_2021, möbus2022optimal,Tim_TrotterZeno_2024}.
 In \cite[Corollary~3.2]{BeckerDattaSalz_Zeno_2021} the spectral condition on $M$ was shown to be equivalent to the mixing limit \eqref{eq:MixingOperationIntro} to hold in uniform topology, i.e. $ \sup_{\|\rho\|_1=1}\|M^n(\rho)-P(\rho)\|_1\xrightarrow[n\to\infty]{}0.$   
	 A first result which ensured existence of the quantum Zeno limit beyond the ubiquitous spectral condition was given in \cite[Theorem 2]{BeckerDattaSalz_Zeno_2021} using a novel pertubation series technique. Here, instead of demanding the mixing limit \eqref{eq:MixingOperationIntro} to hold in uniform topology, it is only assumed to hold in strong toplogy, i.e. pointwise $\lim_{n\to\infty} M^n(\rho) = P(\rho)$ for all states $\rho.$ Demanding merely mixing in strong topology is natural for many physical systems of interest, e.g. photonic systems \cite{BeckerDattaSalz_Zeno_2021,GondolMobusRouze_eEnergyPreservingEvolutions_2024}, as mixing in uniform topology does often not hold in those cases (see Example~\ref{ex:Atten}).  
  
   The mentioned result on the Zeno limit in strong topology, \cite[Theorem 2]{BeckerDattaSalz_Zeno_2021}, is, however, non-quantitative and hence does not provide any bound on the speed of convergence of the respective Zeno limit. Here, we aim to close this gap by deriving an estimate of the convergence speed of the Zeno limit provided one for the respective mixing limit. 

\medskip

As a continuous variant of the quantum Zeno process, we consider the so-called \emph{strong damping limit}~\cite{Zanardi_StrongDamping1_2014,Zanardi_StrongDamping2_2015,Pascazio_NoiseInd_2004,Mac_StrongDampingRand_2016,Burgarth_GeneralisedAdiabic_2019}. In this context, the discrete and periodic implementation of the quantum operation is substituted by an additional term $\cK$ in  the generator of the dynamical semigroup. %Specifically, 
The total dynamics is hence governed %dictated 
by the generator
\begin{align*}
    \cL_{\text{total}}(\gamma)\coloneqq \gamma\cK + \cL,
\end{align*}
with $\gamma\ge 0$ quantifying the strength of interaction of the supplementary generator, $\cK$. Analougosly to \eqref{eq:MixingOperationIntro}, the  dynamics of the additional generator $\left(e^{t\cK}\right)_{t\ge0}$ is assumed to be \emph{mixing}, in the sense that it converges to its fixed point space in the infinite time limit, i.e.~
    \begin{align}
    \label{eq:MixingDymIntro}
        e^{\gamma\cK}\xrightarrow[\gamma\to\infty]{} P
    \end{align}
    in some reasonable topology with $P$ being the projection onto the nullspace of the generator $\cK.$  We then analyse the total dynamics given by the \emph{damped evolution operator} $e^{t\cL_{\text{total}}(\gamma)} = e^{t(\gamma\cK+\cL)}$ in the limit of infinite interaction strength $\gamma\to\infty.$
  This limit process, if it exists, is called the \emph{strong damping limit}. Again, we expect to recover an effective Zeno dynamics on the fixed point space, which precisely means the convergence
    \begin{align}
    \label{eq:StrongDampingIntro}
       e^{t\cL_{\text{total}}(\gamma)}= e^{t(\gamma\cK+\cL)}\xrightarrow[\gamma\to\infty]{} e^{tP\cL P}P.
    \end{align}
    The effect of strong damping is at the core of a proposed paradigm for photonic dynamically error corrected qubits with the capability to perform universal quantum computation \cite{Mirrahimi_DynamicErrorCatQubits_2014,Mirri_New_2019,Mirrahimi_LectureNotes_2023}. Here, the authors consider mixing dynamics $\left(e^{t\cK}\right)_{t\ge 0},$ in the sense of \eqref{eq:MixingDymIntro}, whose fixed point space, with projection $P$, is considered to be the code space of the computation. This dynamics is implemented in a strongly interacting fashion (i.e.~$\gamma$ large in the above) to drive the system into the code space. Typical errors, like photon dephasing errors and single-photon loss, are then corrected due to the effect of strong damping \eqref{eq:StrongDampingIntro}: More precisely, in the language above these errors are modelled by the $\cL$ term in the total generator $\cL_{\text{total}}(\gamma).$
    The generator $\cK$ is then chosen cleverly such that in the strong damping limit we have for the effective dynamics $e^{tP\cL P}P = P$ and hence the information stored in the code space is protected. Moreover, the authors propose different generators $\cL'$ such that for the strong damping limit of $\cL'_{\text{total}}(\gamma) = \gamma\cK +\cL'$ the effective dynamics $e^{tP\cL'P}P$ corresponds to a gate on the code space. In that way they find an universal set of quantum gates on these dynamically error corrected code spaces.

    In both finite \cite{Zanardi_StrongDamping1_2014,Zanardi_StrongDamping2_2015,Burgarth_GeneralisedAdiabic_2019} and infinite dimensions \cite{G,BZ} the strong damping limit \eqref{eq:StrongDampingIntro} has been proven to hold under certain strong assumptions: All of these results relied on the generators $\cK$ and $\cL$ being bounded and the semigroup $\left(e^{t\cK}\right)_{t\ge 0}$ satisfying a spectral condition comparable to the one used for proving quantum Zeno limits in \cite{MobWolf19,BeckerDattaSalz_Zeno_2021,möbus2022optimal}.

\subsection{Summary of main results}

Instead of focusing on the special case of $M$ being a quantum operation on the Banach space of trace class operators $\cT(\cH)$ on some Hilbert space $\cH,$ we allow for any contraction $M$ on some general Banach space $X$, i.e.~with operator norm satisfying $\|M\| = \sup_{\|x\|=1}\|Mx\| \leq 1$. Moreover, for proving the strong damping limit we consider $\cK$ to be a possibly unbounded generator of strongly continuous contraction semigroup $\left(e^{t\cK}\right)_{t\ge0}$ on some Banach space $X$ (see Section~\ref{sec:semigroups} for more details).

We prove quantum Zeno- and strong damping limits in a unified fashion, while going beyond the above-mentioned spectral condition which most of the previous results in the literature relied on. Instead we merely assume that the mixing limits \eqref{eq:MixingOperationIntro} and \eqref{eq:MixingDymIntro} hold in strong topology, i.e. 
\begin{align}
\label{eq:MixDAMPMan}
    \lim_{n\to\infty}M^nx = Px,\quad\quad \lim_{\gamma\to\infty}e^{\gamma \cK}x=Px, 
\end{align}
for all $x\in X$ and some $P\in\cB(X).$ In both cases we find for all $\cL\in\cB(X)$ convergence to an effective dynamics as
\begin{align}
\label{eq:ZENSTRONINTRO}
    \lim_{n\to\infty}\left(Me^{t\cL/n}\right)^n x = e^{tP\cL P}Px, \quad\quad \lim_{\gamma\to\infty}e^{t(\gamma \cK+\cL)}x = e^{tP\cL P}Px,
\end{align}
for all $x\in X$ and $t>0,$ compare Theorems~\ref{thm:StrongZenoGeneralQuant} and~\ref{thm:StrongDamping} for formal statements of the results. Furthermore, both results on Zeno- and strong damping limits, provide bounds on the speed of convergence given control of the convergence speeds of the respective mixing limit in \eqref{eq:MixDAMPMan}. 

Due to this novel bound on the speed of convergence, our result on the quantum Zeno limit, Theorem~\ref{thm:StrongZenoGeneralQuant}, can be seen as an improvement over \cite[Theorem 2]{BeckerDattaSalz_Zeno_2021}, which also applied for quantum operations $M$ satisfying the mixing limit \eqref{eq:MixDAMPMan} in strong topology. Furthermore, Theorem~\ref{thm:StrongDamping} is, to the best of our knowledge, the first result of strong damping with unbounded generators $\cK$ and also without any spectral condition for the semigroup $\left(e^{t\cK}\right)_{t\ge 0}.$

For the unified proof of Zeno- and strong damping limits we consider a novel operator product, which we call the \emph{generalised binomial product}. We analyse its asymptotic behaviour in Theorem~\ref{thm:StrongBinoFormula} by providing a quantitative version of the pertubabition series technique used in \cite[Theorem 2]{BeckerDattaSalz_Zeno_2021}. We then show how both the Zeno product $\left(Me^{t\cL/n}\right)^n$ and the damped evolution operator $e^{t(\gamma\cK+\cL)}$ can be written as such an operator product and conclude their respective convergence for $n\to\infty$ and $\gamma\to\infty$ from the asympotics of the generalised binomial product.

\bigskip

To illustrate our results, we apply them to show Zeno- and strong damping limits for $M$ and $e^{t\cK}$ modelling photon loss of a single-mode bosonic system, i.e. for the so-called \emph{bosonic quantum limited attenutor channel and semigroup}.\footnote{In the following, we often just write attenuator for named channel or semigroup} Importantly, as we see in Example~\ref{ex:Atten} and Section~\ref{sec:Atten} below, the attenuator channel and semigroup are not mixing in uniform topology but only in strong topology. Therefore, this provides an example for which previous methods relying on the above mentioned spectral condition cannot be applied. Moreover, we provide explicit bounds on the speed of convergence of the respective mixing limits. This then enables us to put our above-mentioned results on convergence speeds for Zeno and strong damping limits into use:

\begin{example}[Photon loss - Attenuator channel and semigroup]
\label{ex:Atten}
We consider as Banach space the space of trace class operators, $\cT(\cH)$, on some separable Hilbert space $\cH.$
For $0\neq\eta\in\C$ with $|\eta|<1$ we define the (bosonic quantum limited) attenuator channel $\Phi_\eta^{\text{att}}$ (\cite[Chapter 3.5.3]{P98} and \cite{DePalmaGiovanneti_Attenuator_2016,Winter_EnergyConstrDiamond_2017}) through the relation
\begin{align}
\label{eq:AttCohMain}
    \Phi^{\text{att}}_\eta(\kb{\alpha}) = \kb{\eta\alpha},
\end{align}
where we denoted for $\alpha\in\C$ the coherent state defined by $$\ket{\alpha}= e^{-|\alpha|^2/2} \sum_{m=0}^\infty\frac{\alpha^m}{m!}\ket{m},$$
and $\left(\ket{m}\right)_{m\in\N_0}$ denotes the eigenbasis of the number operator (see Section~\ref{sec:Atten} for details). As shown in Lemma~\ref{lem:AttBasics} below, the relation \eqref{eq:AttCohMain} uniquely defines a quantum channel. The attenuator channel can be considered to model the action of a beam splitter as we outline in Section~\ref{sec:ExamplesZeno}.

Choosing the parametrisation $\eta(t) \coloneqq e^{-t}$ we can define the attenuator semigroup $\left(\Phi^{\text{att}}_{\eta(t)}\right)_{t\ge 0}.$ As discussed around \eqref{eq:AttSemigroupDef}, this indeed defines a strongly continuous semigroup. Therefore, it has a generator $\cK_{\text{att}}$ with dense domain $\cD(\cK_{\text{att}})$ and we shall write $$e^{t\cK_{\text{att}}} = \Phi^{\text{att}}_{\eta(t)}$$ in the following. 

As derived in \eqref{eq:AttStrongPowerConv1} and \eqref{eq:AttStrongMixing} below, the attenuator channel and semigroup are strongly mixing: We have for all states $\rho$
\begin{align}
\label{eq:AttMixExMain}
    \lim_{n\to\infty}\left(\Phi^{\text{att}}_\eta\right)^n(\rho) = P(\rho) \quad\quad\quad\text{and}\quad\quad\quad \lim_{\gamma\to\infty}e^{\gamma\cK_{\text{att}}}(\rho) = P(\rho),
\end{align}
where we have denoted the projection $P(\placeholder) = \Phi^{\text{att}}_0(\placeholder) =  \kb{0}\, \Tr\, (\placeholder).$\footnote{Note however, that these limits do not hold in uniform topology as we can show that
\begin{align}
    \left\|\left(\Phi^{\text{att}}_\eta\right)^n - P\right\| = \left\|e^{\gamma\cK_{\text{att}}} - P\right\| = 2
\end{align}
for all $n\in\N$ and $\gamma\ge 0$ (see \eqref{eq:AttNotUniformCont}). Therefore, using \cite[Corollary 3.2]{BeckerDattaSalz_Zeno_2021} this shows that the attenuator does not satisfy the spectral condition mentioned above on which most of the previous Zeno limit results relied on.}

Due to \eqref{eq:AttMixExMain}, the attenuator channel and semigroup satisfy the assumptions of our Theorem~\ref{thm:StrongZenoGeneralQuant} and Theorem~\ref{thm:StrongDamping} respectively. In particular, we get for all $\cL\in\cB(\cT(\cH))$ the Zeno- and strong damping limits
\begin{align*}
    \lim_{n\to\infty} \left(\Phi^{\text{att}}_\eta e^{t\cL/n}\right)^n (\rho) = e^{tP\cL P}P(\rho) = e^{t\Tr(\cL(\kb{0}))}\, \kb{0}
\end{align*}
and 
\begin{align*}
    \lim_{\gamma\to\infty} e^{t(\gamma \cK_{\text{att}}+\cL )}(\rho) = e^{tP\cL P}P(\rho) =  e^{t\Tr(\cL(\kb{0}))}\, \kb{0} .
\end{align*}

Moreover, we show around \eqref{eq:AttStrongMixingBound} below that for states $\rho$ having finite average particle number $\Tr(N\rho)<\infty$ (see~\eqref{eq:AverageParticleNumber} for definition) the convergences in \eqref{eq:AttMixExMain} in fact even happen exponentially fast, as we have
\begin{equation*}
	\left\|\left(\Phi^{\text{att}}_\eta(\rho) \right)^n- P(\rho)\right\|_1 \le 4|\eta|^n\, \Tr\left((N+\1)\rho\right) \xrightarrow[n\to\infty]{}0
\end{equation*}
and 
\begin{equation}
	\left\|e^{\gamma\cK_{\text{att}}}(\rho) - P(\rho)\right\|_1 \le 4e^{-\gamma}\, \Tr\left((N+\1)\rho\right) \xrightarrow[\gamma\to\infty]{}0.
\end{equation}
From the more detailed version of our results, stated in Theorem~\ref{thm:StrongZenoGeneralQuant} and~\ref{thm:StrongDamping}, %this implies 
 we then infer the following speeds of convergence for the Zeno- and strong damping limit, respectively:
\begin{align}
\left\|\left(\Phi^{\text{att}}_\eta e^{t\cL/n}\right)^n (\rho) -  e^{t\Tr(\cL(\kb{0}))}\,\kb{0}\right\|_1 = \mathcal{O}_{\rho,\cL}\left(\frac{\log(n)}{n}\right)
\end{align}
and
\begin{align}
\left\|e^{t(\gamma \cK_{\text{att}}+\cL )}(\rho) - e^{t\Tr(\cL(\kb{0}))} \, \kb{0}\right\|_1 = \mathcal{O}_{\rho,\cL}\left(\frac{\log(\gamma)}{\gamma}\right).
\end{align}
Here, $\mathcal{O}_{\rho,\cL}$ denotes that the constants corresponding to the Landau-$\mathcal{O}$ notation depend on $\rho$ and $\cL$ and are finite if $\Tr(N\rho)<\infty$ and furthermore $\cL$ does not generate infinite number of particles from the \emph{vacuum state} $\kb{0},$ with precise meaning given in Proposition~\ref{prop:AttZenoDamp} and Remark~\ref{rem:AttZeno}.

\end{example}

\smallsection{Outline of the rest of the paper:}	
    
 \noindent In Section~\ref{chap:Preliminaries} we provide some basic notations and facts which are need for the derivations of the results of this paper.  In Section~\ref{sec:GeneralBino} we define the generalised binomial product and analyse its asymptotic behaviour in Theorem~\ref{thm:StrongBinoFormula}. Then in Section~\ref{sec:ZenoAndStrongFromBino} we show how the quantum Zeno- and strong damping limit \eqref{eq:ZENSTRONINTRO} follow from Theorem~\ref{thm:StrongBinoFormula}.  In Section~\ref{sec:ExamplesZeno} we give a discussion of the speed of convergence for Zeno- and strong damping limits provided in Theorems~\ref{thm:StrongZenoGeneralQuant} and~\ref{thm:StrongDamping}. Here, we give an illustration of the results by considering as examples the case in which the mixing limits~\eqref{eq:MixingOperationIntro} and~\eqref{eq:MixingDymIntro} hold in uniform topology and furthermore the attenuator channel and semigroup.  We then continue to give the proof of Theorem~\ref{thm:StrongBinoFormula} in Section~\ref{sec:ProofOfBino}. The paper is ended with a summary of the results and a discussion of open problems in Section~\ref{sec:Summary}.

\bigskip

 \comment{ \red{We start with reviewing some mathematical preliminaries which are needed for the deriving the results mentioned above. In particular we summarise some well-known results from spectral- and operator ergodic theory as well as the holomorphic functional calculus and semigroup theory on Banach spaces.}}

%\item In Section \ref{sec:channelconstraints}, we prove the convergence of the Zeno product for unbounded generators under an alternative condition.
%	 \item We study Zeno-limits in Hilbert-Schmidt norm in Section \ref{sec:HSnorm}. 

	\section{Preliminaries}
    \label{chap:Preliminaries}
    
    In this section we recall some well-known definitions and facts which are used for the derivations of this paper.
    
    \smallsection{Basic notation:} For $n\in\N$ we denote the set $[n]\coloneqq\{1,\cdots,n\}.$ The set of non-negative real numbers is written as $\R_+.$ In this thesis, if not explicitly stated otherwise, $\log$ denotes the natural logarithm with base $e.$ We furthermore make use of the Landau-$\mathcal{O}$ notation: For $x$ being a real parameter we write $f(x)=\mathcal{O}(g(x))$ as $x\to\infty$ if there exists a constant $C\ge 0$ and $x_0\ge 0$ such that $|f(x)|\le Cg(x)$ for all $x\ge x_0.$ 
    
	\label{sec:Mathprelim}
    \subsection{Operators on Banach spaces}
	In this section we review some basic definitions and facts about operators on complex Banach spaces, all of which can be found in one of the books \cite{dunford_linear_1988,Yos80,ReedSimon_FunctionalAnalysis_1976}. 
 
    Let $X$ be some complex Banach space with norm denoted by $\|x\|$ for $x\in X.$ For a linear operator $T:\cD(T)\to X$ with domain $\cD(T)\subseteq X$ we write \begin{align*}
        \ker(T)&\coloneqq\left\{x\in\cD(T)\,\Big|\,Tx = 0\right\},\\
        \ran(T)&\coloneqq\left\{y\in X\,\Big|\,y=Tx,\,x\in\cD(T) \right\}
    \end{align*} to denote the \emph{kernel} and \emph{range} or \emph{image}  respectively. We say $T$ is \emph{closed} if for all sequences $(x_n)_{n\in\N}\subseteq\cD(T)$ such that $\lim_{n\to\infy}x_n=x$ and $\lim_{n\to\infty}Tx_n = y$ for some $x,y\in X$ we already have $x\in\cD(T)$ and $Tx=y.$ This is equivalent to say that the \emph{graph} of $T$
    \begin{align*}
        \Gamma(T) \coloneqq \left\{(x,Tx)\Big| \,x\in\cD(T)\right\} \subseteq X\times X
    \end{align*} 
    is closed or $\cD(T)$ equipped with the \emph{graph norm} $\|x\|_T \coloneqq \|x\| + \|Tx\|$ is complete.
    
 We denote by $\cB(X)$ the set of bounded linear operators on $X$. $\cB(X)$ is a Banach space itself equipped with the \emph{operator norm} which for $T\in\cB(X)$ is given by
 \begin{align*}
     \|T\| \coloneqq \sup_{\|x\|=1}\left\|Tx\right\|.
 \end{align*}
 We denote by $\id \in \cB(X)$ the identity operator acting on $X.$ We call $T\in\cB(X)$ a \emph{contraction} if $\|T\|\le1.$ 
 
 We say a map 
 \begin{align*}
    A &\to \cB(X)\\
    a&\mapsto T_a,
 \end{align*}
 with $A$ being some topological space, is \emph{uniformly-} or \emph{norm continuous} if $a\mapsto T_a$ is continuous, i.e.~with respect to the operator norm in $\cB(X)$, and \emph{strongly continuous} if for all $x\in X$ the map $a\mapsto T_ax$ is continuous. Similarly, we say a sequence $\left(T_n\right)_{n\in\N}\subseteq\cB(X)$ is \emph{uniformly convergent} if there exists a $T\in\cB(X)$ such that $\lim_{n\to\infty}T_n=T$ and \emph{strongly convergent} if $\lim_{n\to\infty}T_nx=Tx$ for all $x\in X.$
 
    For $T\in\cB(X)$ we say a complex number $\lambda \in {\mathbb{C}}$ is in the {\em{resolvent set}}, $\res(T)$, if $(\lambda \id - T)$ is a bijection. \footnote{More generally we can also define the spectrum of an possibly unbounded operator $T$ with domain $\cD(T)\subseteq X,$ as the set of complex numbers $\lambda \in {\mathbb{C}}$ such that $(\lambda \id - T)$ is a bijection and the inverse operator $(\lambda \id - T)^{-1}$ is bounded. Note that we do not need to require boundedness of the resolvent in the case of $T$ being bounded by the bounded inverse theorem. The spectrum of $T:\cD(T)\to X$ is then defined analogously as $\spec(T)=\C\setminus\res(T).$} 
    For $\lambda \in \res(T)$, the operator $R_\lambda(T) := (\lambda - T)^{-1}\in\cB(X)$ is called the {\em{resolvent}} and is well-defined. Here and henceforth, $(\lambda - T)$ denotes $(\lambda \id - T)$.

 The {\em{spectrum}} of $T$, denoted as $\Spec(T)$, is the complement of the resolvent set and is a closed set in $\C$. The spectrum can be regarded as the set of singularities of the resolvent map $\lambda\mapsto\left(\lambda -T\right)^{-1}.$

	\medskip
	
\comment{ Maybe include $|\lambda|> r(T)$ we have
 \begin{align*}
 \left(\lambda-T\right)^{-1} = \sum_{n=0}^\infy \frac{T^{n}}{\lambda^{n+1}}.
 \end{align*}
 Laurant series.. singularities.. $\|(\lambda-T)^{-1}\|\le \frac{1}{|\lambda|-\|T\|}$ for $|\lambda|>\|T\|.$}
	
\comment{We now recall some general concepts from semigroup theory, which we employ in our proofs.}

\subsection{Fixed points of contractions on Banach spaces}
	\label{sec:FixedPointPrel}
 In this section we review some basic results about fixed point spaces of contractions on a Banach space $X$. In particular we discuss the relevant facts we need from operator ergodic theory.
 
	We say $x\in X$ is a \emph{fixed point} of a contraction\footnote{More generally, all statements in this section remain true if $M$ is not a contraction but only \emph{power bounded}, i.e.~$\sup_{n\in\N}\|M^n\|<\infy$.} $M \in \cB(X)$ if $Mx=x$ and hence the \emph{fixed point space} of $M$ shall be denoted by \begin{align}
	    \F(M)\coloneqq\Big\{x\in X\Big| Mx =x\Big\} = \ker\left(\id-M\right).
	\end{align}
 By boundedness and linearity of $M$ this denotes a closed subspace of $X.$ We can define a projection onto $\cF(M)$ which, however, on a general Banach space might be unbounded. To obtain a bounded projection we may restrict to the so-called \emph{mean ergodic subspace,} defined below. Moreover, this projection is then given by an explicit limit evolving the contraction $M.$
 
For that construction, consider for $n\in\N$ the \emph{Ces\`aro mean}
	\begin{align}
	\label{eq:AverageOp}
	A_n \coloneqq \frac{1}{n}\sum_{k=1}^{n}M^k.
	\end{align}
	The {\em{mean ergodic subspace}} of $M$, which is the subspace of $X$ on which \eqref{eq:AverageOp} has a strong limit, shall be denoted by
\[	X_{\text{me}}(M) = \Big\{x\in X\Big| \,x_{\infy} \coloneqq  \lim_{n\to\infty}A_n x \text{ exists}\Big\}.\]
Using that the operators $A_n$ are uniformly bounded, in fact even contractions, for all $n\in\N,$ it is easy to see that $X_{\text{me}}(M)$ is a closed subspace of $X.$

For $x\in X_{\text{me}}(M)$ we immediately see that $x_{\infty}\in\cF(M),$ which follows by
\begin{align}
\label{eq:CesarToFixed}
    Mx_\infty = \lim_{n\to\infy} \frac{1}{n}\sum_{k=1}^n M^{k+1}x = \lim_{n\to\infy} \frac{1}{n}\sum_{k=1}^n M^{k}x = x_\infty,
\end{align}
where for the second to last equality we have used that $\lim_{n\to\infy} M/n = \lim_{n\to\infty} M^{n+1}/n = 0$ since $M$ is a contraction. This then immediately also implies $A_nx_\infty= x_\infty$ for all $n\in\N$ and hence $x_\infty\in X_{\text{me}}(M).$ Hence, we can define the linear map 
\begin{align*}
    A_{\infty}: X_{\text{me}}(M) &\to X_{\text{me}}(M)\\ x&\mapsto x_{\infty} = \lim_{n\to\infy} A_nx.
\end{align*}
Furthermore, we have already established $\ran(A_\infty) = \cF(M),$ as one inclusion follows by \eqref{eq:CesarToFixed} and the other is immediate by definition.
Note that since $A_nA_\infy = A_\infty,$ we see by definition that $A_\infy$ is a projection, i.e.~$A^2_\infty = A_\infty.$ Moreover, $A_\infy$ is in fact a bounded operator, i.e.~$A_\infy\in\cB(X_{\text{me}}(M)),$ as for $x\in X_{\text{me}}(M)$  we have
\begin{align*}
\left\|A_{\infty}x\right\| \le \lim_{n\to\infy} \frac{1}{n}\sum_{k=1}^n\left\|M^k x\right\| \le \|x\|.
\end{align*}
By the same argument as in \eqref{eq:CesarToFixed} we see that also $M$ leaves its mean ergodic subspace invariant, i.e.~$M X_{\text{me}}(M)\subseteq X_{\text{me}}(M),$ and for all $x\in X_{\text{me}}(M)$ we have $A_{\infty} Mx = A_{\infy}x.$

	{\em{Yosida's Mean Ergodic Theorem}} (cf.~\cite[Chapter 2]{Krengel85} or \cite[Chapter VIII.3.]{Yos80}) gives the following complete characterisation of the mean ergodic subspace:
	\begin{align*} 
	X_{\text{me}}(M) = \F(M) \oplus\overline{\operatorname{ran}\left(\id-M\right)}.
	\end{align*}
    It is easy to see that $A_{\infty}(\overline{\operatorname{ran}\left(\id-M\right)}\, )= \{0\},$ and the above shows that $A_{\infty}$ is the bounded projection onto the fixed point space $\cF(M),$ i.e.~to summarise
    \begin{align*}
    \ran(A_\infty) &= \cF(M),\quad\quad\quad\quad\ker(A_{\infty}) = \overline{\operatorname{ran}\left(\id-M\right)},\\A^2_{\infty} &= A_{\infty},\quad\quad\quad\quad\quad\quad\, A_\infty M =MA_\infty = A_\infty.
    \end{align*}
     We call an operator $M$ \emph{mean ergodic}, if $X_{\text{me}}(M) = X.$ Moreover, if $\lim_{n\to\infy}A_n = A_\infty$ holds true even in operator norm, we say $M$ is \emph{uniformly mean ergodic}. \emph{Lin's uniform ergodic theorem} \cite{Lin_UniformErgodicTheorem_1974} gives full characterisation of uniformly mean ergodic contractions. It states that the following are equivalent for a contraction $M:$
     \begin{enumerate}
         \item $M$ is uniformly mean ergodic, i.e.~$A_n = \frac{1}{n}\sum_{k=1}^{n}M^k$ converges in operator norm as $n\to\infty$.
         \item $\ran(\id -M)$ is closed and $X=\F(M) \oplus\operatorname{ran}\left(\id-M\right).$
         \item $\ran(\id -M)$ is closed.
         \item $\ran(\id -M)^2$ is closed.
     \end{enumerate}
     We say $M$ is \emph{strongly power convergent} if
     \begin{align}
     \label{eq:PrelStrongPower}
         \lim_{n\to\infy} M^nx = Px
     \end{align}
     for all $x\in X$ and some $P\in \cB(X).$
   In that case we clearly also have that $M$ is mean ergodic and $P$ is the projection onto the fixed point space $\cF(M)$, i.e.~$P=A_\infy.$
   
   If $\lim_{n\to\infy}M^n = P$ holds even true in operator norm, we say $M$ is \emph{uniformly power convergent}. In \cite[Corolloray 3.2]{BeckerDattaSalz_Zeno_2021}  gives a characterisation of uniformly power convergent contractions. In particular, we see that if a contraction is uniformly power convergent, it already gives that the convergence $\lim_{n\to\infy}M^n = P$  happens exponentially fast. Moreover, uniform power convergence is also equivalent to a certain spectral gap condition on $M,$ which was at the core of most previous proofs of the quantum Zeno limit.

\subsection{Dynamical semigroups}\label{sec:semigroups}
	In the following we recall some general concepts from semigroup theory (see~\cite{EngNag00} for more details).
	Let $X$ be a Banach space: we say $(T(t))_{t\ge0} \subseteq \cB(X)$ is an \emph{one-parameter semigroup} if
	\begin{enumerate}
	    \item $T(t)T(s) = T(t+s),$ for all $t,s\ge0,$
	    \item $T(0) = \id$.
	\end{enumerate}
A one-parameter semigroup is be {\em{uniformly- or norm continuous}} if it is at $t=0,$ i.e.
$$\lim_{t\downarrow 0}\|T(t) - \id\| = 0.$$ On the other hand, a semigroup is {\em{strongly continuous}} if for all $x\in X$ we find $$\lim_{t\downarrow 0}\|(T(t) - \id)x\| = 0. $$
	For any such semigroup, we can define the densely-defined and closed \cite[Chapter II Theorem 1.4]{EngNag00} generator $\cL$ by
	\begin{align}
	\label{eq:DefGenerator}
	\cL x \coloneqq\frac{d}{d t}\Big|_{t=0} T(t)x \coloneqq\lim_{t\downarrow 0} \frac{T(t) - \id}{t\,}x
	\end{align}
	for all $x$ in the domain $\cD(\cL)\subseteq X$, which is the set of $x$ for which the strong limit on the right hand side of \eqref{eq:DefGenerator} exists. The generator is bounded if and only if the semigroup is uniformly continuous, in which case $$T(t) = e^{t\cL}$$ with right hand side being in terms of the usual exponential series.
	For \emph{contraction semigroups}, i.e.~semigroups satisfying $\|T(t)\|\le 1$ for all $t\ge0$, we can recover the semigroup from its generator as follows:  The spectrum of the generator $\cL$ of a contraction semigroup is contained in the left half plane of $\CC$ and in addition the resolvents satisfy the bound \cite[Chapter II Theorem $3.5$]{EngNag00}
	\begin{equation}
	\label{eq:ResolventBound}
	 \|\lambda\left(\lambda -\cL\right)^{-1}\| \le 1,\quad \text{ for all }\lambda >0.
	\end{equation}
    Hence, for each $s\in (0,\infty)$ we can define the $s^{th}$ \emph{Yosida approximant} of the generator by
	\begin{equation}
	\label{eq:YosidaOperator}
	\cL_s = s\cL\left(s-\cL\right)^{-1},
	\end{equation}
	which are bounded operators satisfying $\|\cL_s\| \le s$ and in addition
	\begin{align*}
	   \cL_sx \xrightarrow[s\to\infty]{} \cL x, \quad\text{ for all }x\in\cD(\cL).
	\end{align*}
         From the Yosida approximants, the semigroup can be recovered as the strong limit
	\[ \lim_{s\to\infty} e^{t\cL_s}x = T(t)x=: e^{t\cL}x \quad \forall \, x \in \cB(X).\]

For $\cK$ with domain $\cD(\cK)\subseteq X$ being some generator of a strongly continuous contraction semigroup and $\cL\in\cB(X)$ being some perturbation, also $\cK + \cL$ with domain $\cD(\cK +\cL) = \cD(\cK)$ generates of a strongly continuous semigroup $\left(e^{t(\cK+\cL)} \right)_{t\ge 0}\subseteq\cB(X)$ \cite[Chapter III Theorem 1.3]{EngNag00}. Moreover, it satisfies the norm bound
\begin{align}
\label{eq:SumGenNormBound}
\left\|e^{t(\cK+\cL)}\right\| \le e^{t\|\cL\|}
\end{align}
for all $t\ge 0$.

We say a contraction semigroup $\left(e^{t\cK}\right)_{t\ge 0}$ is \emph{strongly mixing} if 
\begin{align}
\label{eq:StronglyMixing}
    \lim_{\gamma\to\infty}e^{\gamma\cK}x = Px
\end{align}
for all $x\in X$ and some $P\in\cB(X).$ Similarly to the discussion of strongly power convergent contractions in Section~\ref{sec:FixedPointPrel}, it is easy to see that by \eqref{eq:StronglyMixing} we directly get $e^{t\cK}P= Pe^{t\cK} = P$ for all $t\ge0$ and by that also $P^2 = P.$ More precisely, \eqref{eq:StronglyMixing} gives equality of all fixed point spaces, i.e.~$\cF(e^{t\cK})=\cF(e^{s\cK})$ for all $t,s\ge 0$ and that $P$ is the projection onto that fixed point space $\ran(P)=\cF(e^{t\cK}),$ which equivalently is also given by $\ker(\cK).$\footnote{Note that in general we have $\bigcap_{t\ge 0}\cF(e^{t\cK})= \ker(\cK).$ To see this note first that every $x\in\bigcap_{t\ge 0}\cF(e^{t\cK})$ we clearly have $x\in\cD(\cK)$ and $\cK x=0$ by simply calculating the derivative as in \eqref{eq:DefGenerator}. For the other inclusion note for $x\in\cD(\cK)$ with $\cK x=0$ and $t\ge 0$ we have $e^{t\cK}x = \int_0^t\frac{d}{ds} e^{s\cK}x + x=\int_0^te^{s\cK}\cK x + x = x.$ } Furthermore, we say the semigroup $\left(e^{t\cK}\right)_{t\ge 0}$ is \emph{uniformly mixing} if the convergence in \eqref{eq:StronglyMixing} does not only happen but also in uniform topology, i.e.~$\lim_{\gamma\to\infty}e^{\gamma\cK} = P$ in operator norm. In \cite[Corollary 4.1.2]{Salzmann_PhDthesis_2023} it is shown that in this case the convergence, in fact, happens already exponentially fast.

\subsection{Operators, states and channels on Hilbert spaces}
\label{sec:OperatorsOnHilbertSpaces}
For some separable and complex Hilbert space $\cH$ we consider certain Banach spaces  consisting of operators on $\cH$. To be able to distinguish the identity on $\cH$ with the one on the corresponding Banach space, we denote it by $\1$ instead of $\id.$ 

We denote the \emph{support} of a linear operator $x\in\cB(\cH)$ by $\supp(x) \coloneqq \ker(x)^\perp.$ 

 A Banach space which is of particular importance to us is $\cT(\cH),$ the space of \emph{trace class operators} on $\cH,$ which is defined to be the set of $x\in\cB(\cH)$ such that $\Tr|x|<\infty.$  It is in fact a Banach space equipped with the \emph{trace norm}
 \begin{align*}
     \left\|x\right\|_1 \coloneqq \Tr|x| = \Tr\sqrt{x^*x}.
 \end{align*}

We say $\rho$ is a \emph{quantum state} (or often simply state in the following) if $\rho\ge 0$ and it has unit trace $\Tr(\rho) =1.$ 

We say a state $\rho$ is \emph{pure} if it is a rank-1 projection and can hence be written as $\rho=\kb{\psi}$ for some normalised $\psi\in\cH.$ 

For $\rho$ and $\sigma$ being states, we denote the \emph{trace distance} by $\frac{1}{2}\left\|\rho-\sigma\right\|_1\in[0,1].$ It can be written as the variational expression 
\begin{align*}
    \frac{1}{2}\left\|\rho-\sigma\right\|_1 = \sup_{\Lambda\in\cB(\cH),\,0\le\Lambda\le\1} \Tr\left(\Lambda\left(\rho-\sigma\right)\right) = \Tr\left(\pi_+\left(\rho-\sigma\right)\right),
\end{align*}
where we denoted the orthogonal projection onto the support of $\rho-\sigma$ by $\pi_+.$ Moreover, for two pure states $\kb{\psi}$ and $\kb{\varphi}$ the trace distance has the form
\begin{align}
\label{eq:PureTraceDistance}
    \frac{1}{2}\left\|\kb{\phi}-\kb{\psi}\right\|_1 = \sqrt{1-|\langle\psi,\varphi\rangle|^2}.
\end{align}

We say an operator $\cN\in\cB(\cT(\cH))$ is \emph{positive}, if \begin{align*}
    \cN(\rho)\ge0,\quad\text{ for all }\rho\in\cT(\cH),\text{ with }\rho\ge 0
\end{align*}
Any positive linear map $\cN$ is in particular hermiticity preserving, i.e.~$\cN(x^*) =\cN(x)^*$ for all $x\in\cT(\cH).$
Furthermore, $\cN\in\cB(\cT(\cH))$ is called \emph{completely positive} if $\left(\cN\otimes\id_d\right)\in\cB(\cT(\cH\otimes\C^d))$ is positive for all $d\in\N$, 
	where we have denoted the identity map on the $d$-dimensional complex square matrices $\cB(\C^d)=\C^{d\times d}$ by $\id_d$.
 
 $\cN\in\cB(\cT(\cH))$ is \emph{trace-preserving }if for all $x\in\cT(\cH)$ we have $\tr(\cN(x)) = \tr(x)$ and \emph{trace non-increasing} if $\tr(\cN(x)) \le \tr(x)$ for all $x\ge 0$. 
 
 We call a completely positive map $\cN\in\cB(\cT(\cH))$ a {\em{quantum operation}} if it is trace non-increasing, and a {\em{quantum channel}} if it is trace-preserving. The set of quantum channels on $\cH$, or often also simply channels for simplicity, is denoted by ${\rm{CPTP}}(\cH).$

    Every completely positive map $\cN\in\cB(\cT(\cH))$ can by \emph{Stinespring's dilation theorem} be written in the form
    \begin{align}
    \label{eq:StindespringComp}
         \cN(x) = \Tr_E\left(VxV^*\right),
     \end{align}
     for some linear bounded operator $V:\cH\to\cH\otimes\cH_E,$ $\cH_E$ being some \emph{environment Hilbert space} and $\Tr_E$ denoting the \emph{partial trace} on $\cH_E$. Moreover, $\cN$ is a quantum operation if and only if $V^*V\le \1$ and a quantum channel if and only if $V^*V =\1.$ In the latter case $V$ is therefore an isometry, which we henceforth call the \emph{Stinespring isometry} of $\cN.$ 
      Equivalently to \eqref{eq:StindespringComp}  a completely positive map $\cN\in\cB(\cT(\cH))$ can  be written in \emph{Kraus decomposition} as
     \begin{align}
     \label{eq:KrausDecomCompPos}
    \cN(x) = \sum_{i=1}^{r_\cN} K_ixK^*_i,
     \end{align}
     where $r_\cN\in\N\cup\{\infty\}$ is called the  \emph{Kraus rank} of $\cN$ and $\left(K_i\right)_{i=1}^{r_\cN}\subseteq\cB(\cH)$ are the so-called \emph{Kraus operators} of $\cN.$ Note, for $r_\cN =\infty$ we have convergence of the series $\sum_{i=1}^{\infty} K^*_iK_i$ in strong topology which implies convergence of the series in \eqref{eq:KrausDecomCompPos} in trace norm. Moreover, $\cN$ is a quantum operation if and only if $\sum_{i=1}^{r_{\cN}} K^*_iK_i\le \1$ and a quantum channel if and only if $\sum_{i=1}^{r_{\cN}} K^*_iK_i = \1$ 
     
   For $\cN\in\cB(\cT(\cH))$ we can denote the \emph{diamond norm} or \emph{completely bounded trace norm} by
   \begin{align*}
       \left\|\cN\right\|_\diamond \coloneqq \sup_{d\in\N, x\in\cT(\cH\otimes\C^d)\, \|x\|_1=1}\left\|(\cN\otimes\id_d)(x)\right\|_1.
   \end{align*}
   Note by the above we can easily see that any quantum operation $\cN$ satisfies $\left\|\cN\right\|_\diamond \le1.$ In particular $\cN\in\cB(\cT(\cH))$ is a contraction as it has operator norm \\$\|\cN\| = \sup_{x\in\cT(\cH)\,\|x\|_1=1}\|\cN(x)\|_1\le1.$
  
    Lastly, using the terminology of Section~\ref{sec:semigroups},
    we call an one-parameter semigroup of quantum channels $\left(\cN(t)\right)_{t\ge 0}$ a \emph{quantum dynamical semigroup.}

\section{Generalised binomial product}
\label{sec:GeneralBino}
In this section we consider an operator generalisation of the product $\left(1+\frac{\lambda}{n}\right)^n,$ where $\lambda\in\C,$ and prove convergence of named generalisation reminiscent of the well-known limit expression of exponential function
\begin{align}
\label{eq:LimitExpFunc}
\lim_{n\to\infty}\left(1+\frac{\lambda}{n}\right)^n = e^\lambda.
\end{align}
 For that, let $R\subseteq \R$  be some compact interval and  $\left(\cL_{n,r}\right)_{n\in\N, r\in R}$ be a family of bounded operators. We assume that, uniformly in $r\in R$, the sequence $(\cL_{n,r})_{n\in\N}$ converges in norm to some operator $\cL_r\in\cB(X)$. In other words we can denote some (abstract) speed of convergence $(s'_n)_{n\in\N}\subseteq \R_+$ such that
	\begin{align}
 \label{eq:LConv}
  \left\|\cL_{n,r}-\cL_r\right\| \le s'_n
	\end{align}
	and $\lim_{n\to\infty}s_n' =0.$ We additionally assume that the map $r\mapsto \cL_r$ is strongly continuous.

 Furthermore, let $R'\subseteq R$ and let $\left(M_r\right)_{r\in R'}\subseteq \cB(X)$ be a family of contractions which is, uniformly in $r\in R'$, strong power convergent, i.e.~for all $x\in X$ we have
 \begin{align}
 \label{eq:Mrstrongpowerconvergenct}
     \lim_{n\to\infty} M^n_rx = Px,
 \end{align}
 for some $P\in\cB(X).$ As discussed in Section~\ref{sec:FixedPointPrel}, this already gives that $P$ is the projection onto the fixed point space $\cF(M_r)$ and, in particular equality of all fixed point spaces, i.e.~$\cF(M_r)=\cF(M_{r'})$ for all $r,r'\in R'$. Defining the sequence $(s_n(x))_{n\in\N}\subseteq \R_+$ by 
 \begin{align}
  \label{eq:Mpowerspeed}
   s_n(x) \coloneqq\sup_{n'\ge n,r\in R'}\|(M^{n'}_r-P)x\|,
 \end{align}
 we clearly have 
	\begin{align}
		\left\|(M_r^n-P)x\right\| \le s_n(x),
	\end{align}
	  and $\lim_{n\to\infty} s_n(x)=0$. Moreover, the map $n\mapsto s_n(x)$ is by definition monotonically decreasing and $x\mapsto s_n(x)$ is continuous.\footnote{To see continuity of $x\mapsto s_n(x)=\sup_{n'\ge n,r\in R'}\|(M^{n'}_r-P)x\| $ use for $x,y\in X$ with $s_n(x) \ge s_n(y)$ that $s_n(x)-s_n(y)\le \sup_{n'\ge n,r\in R'}\left(\|(M^{n'}_r-P)x\| - \|(M^{n'}_r-P)y\| \right)\le 2\|x-y\|.$ }

   For such families $\left(\cL_{n,r}\right)_{n\in\N, r\in R}$ and $\left(M_r\right)_{r\in R'},$ we can now introduce the \emph{generalised binomial product}
   \begin{align}
   \label{eq:GenOpProd}
       \left(M_r + \frac{\cL_{n,r}}{n}\right)^n
   \end{align}
   and analyse its limit in the strong topology as $n$ tends to infinity. In Theorem~\ref{thm:StrongBinoFormula} below, we prove that this limit is given by 
   \begin{align}
   \label{eq:GenOpLimi}
      \lim_{n\to\infy} \left(M_r + \frac{\cL_{n,r}}{n}\right)^nx = e^{P\cL_rP}Px,
   \end{align}
   for all $x\in X.$
    Moreover, in \eqref{eq:SpeedBino} we provide a bound on the speed of convergence in terms of the convergence speeds $s_n'$ and $s_n(x)$ defined in \eqref{eq:LConv} and \eqref{eq:Mpowerspeed}. 
     
   To give some intuition, let us try to connect \eqref{eq:GenOpLimi} back to \eqref{eq:LimitExpFunc}.
   Consider the special case of $M_r =P$ and $\cL_{n,r} = \cL$ for some projection $P\in \cB(X)$ and some operator $\cL\in\cB(X).$ In that case \eqref{eq:GenOpLimi} can be seen to be a direct consequence of \eqref{eq:LimitExpFunc}. To convince ourselves, note first that  
   \begin{align}
    \label{eq:BinoWithProj}
       \left(P + \frac{\cL}{n}\right)^n =  \left(P + \frac{P\cL P}{n}\right)^n +\,\mathcal{O}\left(\frac{1}{n}\right), 
   \end{align}
   which can be seen by using $(\id-P)\left(P + \frac{\cL}{n}\right) = \mathcal{O}(1/n)$, $\left(P + \frac{\cL}{n}\right)(\id-P) = \mathcal{O}(1/n)$ and $\left(P + \frac{\cL}{n}\right)(\id-P)\left(P + \frac{\cL}{n}\right) = \mathcal{O}(1/n^2)$ and hence the sum of all error terms in \eqref{eq:BinoWithProj} is of order $\mathcal{O}(1/n).$
    We can now employ the functional calculus for the operator $P\cL P$ on the Banach space $\ran(P) = \ker(\id-P)$ together with the sequence of functions $f_n(\lambda)=\left(1 +\frac{\lambda}{n}\right)^n,$ for which $f_n(P\cL P) = \left(P + \frac{P\cL P}{n}\right)^n,$ giving in operator norm
   \begin{align}
      \lim_{n\to\infy} \left(P + \frac{\cL}{n}\right)^n = e^{P\cL P}P,
   \end{align}
by using the continuity of the holomorphic functional calculus.\footnote{Note that the convergence $\lim_{n\to\infty} \left(1 +\frac{\lambda}{n}\right)^n = e^\lambda$ is uniform over $\lambda\in K\subseteq\C$ for $K$ compact.}

   The reason for considering the generalised binomial product \eqref{eq:GenOpProd} is by that both the Zeno product, $\left(Me^{t\cL/n}\right)^n$, as well as the damped evolution operator, $e^{t(\gamma\cK+\cL)},$ can be treated in a unified way as they can both be written in the form of \eqref{eq:GenOpProd}. Hence, in Theorem~\ref{thm:StrongZenoGeneralQuant} and Theorem~\ref{thm:StrongDamping} below, we see how the Zeno- and strong damping limit, respectively, can be deduced from Theorem~\ref{thm:StrongBinoFormula}. Here, for the Zeno product, the $r$ dependence of $\cL_{n,r}$ and $M_r$ is trivial, i.e.~$|R|=1$, whereas a non-trivial $r$ dependence becomes necessary to also cover the strong damping limit.

\begin{theorem}[Asymptotics of the generalised binomial product]
	\label{thm:StrongBinoFormula}
Let $R$ be a compact interval and $(\cL_{n,r})_{n\in\N}\subseteq\cB(X)$ be operators which, uniformly in $r\in R$, converge to some operator $\cL_r$. Furthermore, assume that $r\mapsto \cL_r$ is strongly continuous.\\
Let $R'\subseteq R$ and $\left(M_r\right)_{r\in R'}\subseteq \cB(X)$ be contractions such that for all $x\in X$ 
 \begin{align}
 \label{eq:MrstrongpowerconvergenctThm}
     \lim_{n\to\infty} M^n_rx = Px,
 \end{align}  uniformly in $r\in R'$ and for some $P\in\cB(X).$ 
Then
\begin{align}
\lim_{n\to\infty}\left(M_r + \frac{\cL_{n,r}}{n}\right)^n x = e^{P\cL_r P}Px
\end{align}
uniformly in $r\in R'$. Moreover, using the notation established in \eqref{eq:LConv} and \eqref{eq:Mpowerspeed}, we have 
\begin{equation}
\label{eq:SpeedBino}
    \left\|\left(\left(M_r + \frac{\cL_{n,r}}{n}\right)^n - e^{P\cL_r P}P\right)x\right\|\le C\Big(s'_n\|x\| + s^{\sup}_n(x,\cL_r,\mathbf{N})\Big).
\end{equation}
for all $r\in R'$ and $\mathbf{N}=(N_l)_{l\in\N}\subseteq \N$ and some $C\ge 1$ independent of $n$ and $\mathbf{N}.$ Here, we have defined
 \begin{equation}
 \label{eq:ssupBino}
s^{\sup}_n(x,\cL_r,\mathbf{N})\coloneqq\sup_{l\in\N}\left(\left(\sup_{\tilde r\in R}\|\cL_{\tilde r}\|\right)^{-l+1}s_{N_l}\left((\cL_rP)^{l-1}x\right) +\frac{N_l}{n}\|x\|\right).
 \end{equation}
 \end{theorem}
 \begin{remark}
 \label{rem:SpeedOfConv}
    Although the particular form of the bound on the speed of convergence\\ $s^{\sup}_n(x,\cL_r,\mathbf{N})$ in \eqref{eq:SpeedBino} is rather involved, it can be shown to be well-behaved for some interesting examples - see Section~\ref{sec:ExamplesZeno} below where the respective Zeno and strong damping limits are discussed. Here, the idea is that if we have, for a given example, explicit control of the $s_n'$ and $s_n(x)$ defined in \eqref{eq:LConv} and \eqref{eq:Mpowerspeed}, this leads to an explicit bound on the speed of convergence for the generalised binomial product.
    
    Note that by the definition of $s^{\sup}_n(x,\cL_r,\mathbf{N}),$ there is a trade-off between the sequence $\mathbf{N} = \left(N_l\right)_{l\in\N},$ which can be choosen freely, and $n.$ Hence, the idea is to choose an $n$-dependent $\mathbf{N}(n)= \left(N_l(n)\right)_{l\in\N}$ in such a way that both summands in \eqref{eq:ssupBino} decay as $n\to\infty$, i.e.
    \begin{align*}
        \lim_{n\to\infty}s_{N_l(n)}\left((\cL_rP)^{l-1}x\right) = 0 \quad\quad\text{and}\quad\quad\lim_{n\to\infty}\frac{N_l(n)}{n}=0,
    \end{align*}
    and we furthermore have good control on the respective speeds of convergence to get \\$\lim_{n\to\infty}s^{\sup}_n(x,\cL_r,\mathbf{N}(n))=0.$
    
    Since the convergence \eqref{eq:MrstrongpowerconvergenctThm} is assumed to happen in strong topology, and not in uniform topology, note that due to the supremum over $l\in\N$ in \eqref{eq:ssupBino}  $\lim_{n\to\infty}s^{\sup}_n(x,\cL_r,\mathbf{N}(n))=0 $  might not be satisfied in general. In that case Theorem~\ref{thm:StrongBinoFormula} still gives convergence of the generalised binomial product, but without an explicit (and decaying) bound on the speed of convergence. 
 \end{remark}
 We defer the proof of Theorem~\ref{thm:StrongBinoFormula} to Section~\ref{sec:ProofOfBino}. First we state the respective convergence results for the Zeno- and strong damping limits in the next section and see how they are both consequences of Theorem~\ref{thm:StrongBinoFormula}.

\section{Quantum Zeno- and strong damping limits from Theorem~\ref{thm:StrongBinoFormula}}
\label{sec:ZenoAndStrongFromBino}
Using Theorem~\ref{thm:StrongBinoFormula} we can prove convergence of both the Zeno product and the damped evolution operator. This is done in the following section. In both cases the idea is to rewrite the Zeno product, $\left(Me^{t\cL/n}\right)^n$, or the damped evolution operator, $e^{t(\gamma\cK+\cL)},$ as a generalised binomial product \eqref{eq:GenOpProd} by choosing appropriate families of operators $\left(\cL_{n,r}\right)_{n\in\N,r\in R}$ and $\left(M_r\right)_{r\in R'}.$

\subsubsection{Quantum Zeno limit}

   \noindent For the Zeno product, we typically only consider the case $R'=R$ with $|R|=1$ and hence a single contraction $M\in\cB(X).$ In that case, using the notation of \eqref{eq:Mpowerspeed}, we simply have
    \begin{align}
  \label{eq:MpowerspeedZeno}
   s_n(x) =\sup_{n'\ge n}\|(M^{n'}-P)x\|.
 \end{align}
We choose $\cL_n$ to be the \emph{discrete derivative} of the dynamics generated by $\cL$, i.e.
\begin{align}
    \cL_n \coloneqq n\left(e^{t\cL/n}-\id\right).
\end{align}
By that choice, the Zeno product can be written as a generalised binomial product of $M$ and $M\cL_n$ as outlined in the proof of Theorem~\ref{thm:StrongZenoGeneralQuant} below. Moreover, as we show in there, using the notation of \eqref{eq:LConv}, for that particular choice we have good control of the bound of the speed of convergence as $s_n' = \mathcal{O}(1/n).$ This leads to a simplified bound on the speed of convergence for the Zeno limit compared to the one presented in \eqref{eq:SpeedBino} in Theorem~\ref{thm:StrongZenoGeneralQuant}.
\begin{theorem}[Quantum Zeno limit]
\label{thm:StrongZenoGeneralQuant}
Let $M\in\cB(X)$ be a contraction which satisfies the following: For all $x\in X$
\begin{align}
\label{eq:StrongPowerConv}
\lim_{n\to\infty}M^n x =Px,
\end{align}
for some operator $P\in\cB(X)$. Then for all $\cL\in\cB(X)$ and $t\ge 0$ we have
\begin{align}
\label{eq:StrongZenoConv}
\lim_{n\to\infty}\left(Me^{t\cL/n}\right)^n x = e^{tP\cL P}Px.
\end{align}
  Moreover, using the notation established in \eqref{eq:MpowerspeedZeno}, we have 
\begin{equation}
\label{eq:SpeedZeno}    \left\|\left(\left(Me^{t\cL/n}\right)^n - e^{tP\cL P}P\right)x\right\|\le C\,s^{\sup}_n(x,t\cL,\mathbf{N}),
\end{equation}
for all $\mathbf{N}=(N_l)_{l\in\N}\subseteq \N$ and some $C\ge 1$ independent of $n$ and $\mathbf{N}.$ Here, we have denoted
 \begin{equation}
s^{\sup}_n(x,t\cL,\mathbf{N})\coloneqq\sup_{l\in\N}\left(\|t\cL\|^{-l+1}s_{N_l}\left((t\cL P)^{l-1}x\right) +\frac{N_l}{n}\|x\|\right).
 \end{equation}
\end{theorem}
\begin{rem}
For the special case in which $X$ is the space $\cT(\cH)$ of trace class operators over some Hilbert space $\cH$, it is known \cite{A81} that for $x\in\cT(\cH)$ and sequences $\left(x_n\right)_{n\in\N}\subseteq\cT(\cH)$ we have $\| x_n - x \|_1 \xrightarrow[n \to \infty]{} 0$ if and only if $\| x_n \|_1 \rightarrow \| x \|_1$ and $x_n$ is weakly convergent to $x.$ Therefore, often (e.g.~when $M$, or $e^{t\cK}$ in Theorem~\ref{thm:StrongDamping} below, are quantum channels) it is enough to just assume a \emph{weak power-convergence} or \emph{weak mixing}\footnote{With that we mean that the convergences $M^n\xrightarrow[n\to\infty]{}P$ and $e^{\gamma\cK}\xrightarrow[\gamma\to\infty]{}P$ are only assumed to happen with respect to the weak topology on $\cT(\cH)$. This means concretely $\lim_{n\to\infy}\Tr(BM^n(x))=\Tr(BP(x))$ and $\lim_{\gamma\to\infy}\Tr(Be^{\gamma\cK}(x))=\Tr(BP(x))$ for all $x\in\cT(\cH)$ and $B\in\cB(\cH)$.} in Theorems~\ref{thm:StrongZenoGeneralQuant} and~\ref{thm:StrongDamping}.
\end{rem}
\begin{proof}[Proof of Theorem~\ref{thm:StrongZenoGeneralQuant}]
In the following we omit the $t\ge 0$ from the notation of Theorem~\ref{thm:StrongDamping} by redefining $\cL$ to be $t\cL$. We define for $n\in\N$
\begin{align}
    \cL_n \coloneqq n\left(e^{\cL/n}-\1\right).
\end{align}
Note that we can write 
\begin{align*}
    n\left(e^{\cL/n}-\1\right) - \cL = n\int_0^1\frac{d}{ds}e^{s\cL/n}\,ds-\cL = \int_0^1\left(e^{s\cL/n}-\1\right)\cL\,ds=\frac{1}{n}\int_0^1\int_0^1e^{s's\cL/n}\cL^2\,ds'ds,
\end{align*}
which gives
\begin{align}
\label{eq:ZenoSpeedBound}
    \left\|\cL_n-\cL\right\| \le\frac{1}{n} \sup_{s\in[0,1/n]}\left\|e^{s\cL}\right\|\left\|\cL^2\right\|.
\end{align}
With that we can write the Zeno product as 
\begin{equation}
    \left(Me^{\cL/n}\right)^n = \left(M+\frac{M\cL_n}{n}\right)^n.
\end{equation}
The statement \eqref{eq:StrongZenoConv} then directly follows from Theorem~\ref{thm:StrongBinoFormula} giving for $x\in X$
\begin{align*}
    \lim_{n\to\infty}\left(Me^{\cL/n}\right)^nx =e^{PM\cL P}Px =e^{P\cL P}Px,
\end{align*}
where we have used the fact that, due to 
%that necessarily by 
the strong power convergence \eqref{eq:StrongPowerConv}, we have $PM = MP = P.$

The bound on the speed of convergence \eqref{eq:SpeedZeno} is immediate from \eqref{eq:SpeedBino} in Theorem~\ref{thm:StrongBinoFormula} after using that by \eqref{eq:ZenoSpeedBound}  we have $$s'_n\|x\|\le \frac{C\|x\|}{n}\le C s^{\sup}_n(x,\cL,\mathbf{N})$$ for some $C\ge 0.$ 
\end{proof}

\subsubsection{Strong damping limit}

   \noindent Let $\cK \in \cB(X)$ with domain $\cD(\cK)\subseteq X$ being the generator of a strongly continuous contraction group, $\left(e^{t\cK}\right)_{t\ge 0}$, which is strongly mixing as defined and discussed around \eqref{eq:StronglyMixing}, i.e.~
   \begin{align}
	\lim_{\gamma\to\infty} e^{\gamma \cK}x = Px
	\end{align}
	for all $x\in X$ and some $P\in \cB(X)$. 
 
 Again, similarly to the definition of $s_n(x)$ in \eqref{eq:Mrstrongpowerconvergenct}, we define some \emph{(abstract) speed of convergence} $(v_\gamma(x))_{\gamma\ge 0}\subseteq\R_+$ by
 \begin{align}
 \label{eq:Kspeed}
v_{\gamma}(x) := \sup_{\gamma'\ge\gamma}\left\|\left(e^{\gamma' \cK} -P\right)x\right\|,
\end{align}
which clearly satisfies
\begin{align}
\left\|\left(e^{\gamma \cK} -P\right)x\right\|\le  v_{\gamma}(x)
\end{align}
and $\lim_{\gamma\to\infty}v_\gamma(x) = 0.$ Moreover, we have by definition that the map $\gamma\mapsto v_\gamma(x)$ is monotonically decreasing and $x\mapsto v_\gamma(x)$ is continuous.

For $\cL\in\cB(X)$ and $\gamma\ge 0$ we want to analyse the perturbed dynamics generated by 
\begin{align}
\label{eq:TotalDampGenerator}
    \gamma \cK + \cL\quad\quad\text{with}\quad\quad\cD(\gamma \cK + \cL) = \cD(\cK),
\end{align}
in the strong interaction limit $\gamma\to\infty.$ Note, as discussed in Section~\ref{sec:semigroups}, the operator \eqref{eq:TotalDampGenerator} in fact generates a strongly continuous semigroup $\left(e^{t(\gamma\cK+\cL)}\right)_{t\ge 0}.$ In the proof of Theorem~\ref{thm:StrongDamping} below, %the main idea to 
the key step for understanding the behaviour for large $\gamma$ is to rewrite $e^{t(nr\cK+\cL)}$ for $n\in\N$ and  $r\in R'\coloneqq[1,2]\subseteq [0,2]=:R$ as a generalised binomial product \eqref{eq:GenOpProd} by choosing appropriate families of operators $\left(\cL_{n,r}\right)_{n\in\N,r\in R}$ and $\left(M_r\right)_{r\in R'}.$ In particular, we define 
\begin{align*}
    \cL_{n,r} \coloneqq n\left(e^{t(r\cK+\cL/n)} - e^{tr\cK}\right),
\end{align*}
as well as the contractions $M_r = e^{tr\cK}$, and find 
\begin{align*}
    e^{t(nr\cK + \cL)} &= \left(e^{t(r\cK+\cL/n)}\right)^n = \left(M_r + \frac{\cL_{n,r}}{n}\right)^n.
\end{align*}
We can then use Theorem~\ref{thm:StrongBinoFormula} to prove strong convergence in the limit $n\to\infty.$ The freedom in choosing the parameter $r$ and the fact that the convergence in the result of Theorem~\ref{thm:StrongBinoFormula} happens uniformly in $r$, then also enables us to cover the more general convergence as $\gamma\to\infty.$

Moreover, as discussed in more detail in the proof of Theorem~\ref{thm:StrongDamping} below, we see that $\cL_{n,r}$ converges in operator norm to some $\cL_r\in\cB(X)$ which satisfies $P\cL_rP = tP\cL P.$ Using the notation of \eqref{eq:LConv}, we show that we have good control on the corresponding speed of convergence as $s_n' = \mathcal{O}(1/n).$  Similarly as for the quantum Zeno limit discussed in Theorem~\ref{thm:StrongZenoGeneralQuant}, this leads to a simplified bound on the speed of convergence for the strong damping limit compared to the one presented in \eqref{eq:SpeedBino} in Theorem~\ref{thm:StrongZenoGeneralQuant}.

We are now ready to state and prove the result on the strong damping limit, outlined above, in the following theorem.

\begin{theorem}[Strong damping limit]
	\label{thm:StrongDamping}
	Let $\cK$ with domain $\cD(\cK)\subseteq X$ be the generator of a strongly continuous contraction semigroup which satisfies for all $x\in X$
	\begin{align}
	\label{eq:StrongErgodicity}
	\lim_{\gamma\to\infty} e^{\gamma \cK}x = Px,
	\end{align}
	for some $P\in\cB(X).$ Furthermore, let $\cL\in\cB(X)$. Then for all $t >0$
	\begin{align}
	\label{eq:StrongDampingConv}
	\lim_{\gamma\to\infty}e^{t(\gamma \cK+\cL)}x = e^{tP\cL P}Px.
	\end{align}
 Moreover, using the notation in \eqref{eq:Kspeed} we have
 \begin{align}
 \label{eq:StrongDampSpeed}
\left\|\left(e^{t(\gamma\cK + \cL)} -  e^{tP\cL P}P\right)x\right\| \le C v^{\sup}_\gamma(x,t\cL,\mathbf{N}).
\end{align}
for all $\mathbf{N}=\left(N_l\right)_{l\in\N}$ and some $C\ge 1$ independent of $\gamma$ and $\mathbf{N}.$ Here, we have denoted
\begin{align}
  v^{\sup}_\gamma(x,t\cL,\mathbf{N}) \coloneqq  \sup_{l\in\N}\left(\|t\cL\|^{-l+1}v_{N_l}\left((t\cL P)^{l-1}x\right) +\frac{N_l}{\floor{\gamma}}\|x\|\right). 
\end{align}
\end{theorem}

\begin{proof}[Proof of Theorem~\ref{thm:StrongDamping}]
In the following we will omit the $t> 0$ from the notation of Theorem~\ref{thm:StrongDamping} by redefining $\cK$ and $\cL$ to be $t\cK$ and $t\cL$ respectively.
Let $r\in R :=[0, 2]$ and $n\in\N$ and define
\begin{align*}
    \cL_{n,r} \coloneqq n\left(e^{r\cK+\cL/n} - e^{r\cK}\right).
\end{align*}
Note that for $x\in\cD(\cK)$ we can write 
	\begin{align}
	\label{eq:DiffonDomain}
	\left(e^{r\cK+\cL/n} - e^{r\cK}\right)x = \int_0^1 \frac{d}{ds}\left(e^{s(r\cK+\cL/n)}e^{(1-s)r\cK}\right)x\, ds = \frac{1}{n}\int_0^1 e^{s(r\cK+\cL/n)}\cL e^{(1-s)r\cK}x\,ds.
	\end{align}
	Note that we can bound 
	\begin{align}
	\label{eq:NormBound}
\left\|e^{s(r\cK+\cL/n)}\right\| \le e^{s\|\cL\|/n},\quad\quad \left\|e^{(1-s)r\cK}\right\| \le 1
	\end{align} for all $s\in[0,1]$ and $r\in R$ (c.f. \eqref{eq:SumGenNormBound}). Therefore, we see that both operators on the left-hand and right-hand side of \eqref{eq:DiffonDomain} are bounded. Furthermore, using that $\cD(\cK)$ is dense in $X$ (c.f. Section~\ref{sec:semigroups}), we have 
	\begin{align}
	\label{eq:Diff}
	\cL_{n,r} =  \int_0^1 e^{s(r\cK+\cL/n)}\cL e^{(1-s)r\cK}\,ds.
	\end{align}
    Define now the family of bounded operators
    \begin{align*}
        \cL_r \coloneqq\int_0^1 e^{sr\cK}\cL e^{(1-s)r\cK}\,ds.
    \end{align*}
   Using the fact that $\cK$ generates a strongly continuous semigroup, together with the dominated convergence theorem, we see that the map $r\mapsto \cL_r$ is strongly continuous.
    Moreover, we have
    \begin{align*}
       \cL_{n,r}-\cL_r &= \int_0^1 \left(e^{s(r\cK+\cL/n)}-e^{sr\cK}\right)\cL e^{(1-s)r\cK}ds \\& =  \frac{1}{n}\int_0^1 \int_0^1se^{s's(r\cK+\cL/n)}\cL e^{(1-s')sr\cK}\cL e^{(1-s)r\cK}\,ds'ds.
    \end{align*}
    Therefore, using  \eqref{eq:NormBound} again, we have
    \begin{align}
    \label{eq:LSpeedDamp}
        \left\|\cL_{n,r}-\cL_r\right\| \le\frac{1}{n} \sup_{s\in[0,1/n]}e^{s\|\cL\|}\|\cL\|^2.
    \end{align}
 Defining the contraction $M_r = e^{r\cK}$ we can write
\begin{align}
\label{eq:Expansion}
e^{nr\cK + \cL} &= \left(e^{r\cK+\cL/n}\right)^n = \left(M_r + \frac{\cL_{n,r}}{n}\right)^n.
\end{align}
Let $x\in X$ and $R' =[1,2] \subseteq R$ we see by \eqref{eq:StrongErgodicity} that for $r\in R'$ we have 
\begin{align*}
\sup_{r\in[1,2]}\left\|\left(M^n_r  - P\right)x\right\|= \sup_{r\in[1,2]}\left\|e^{n(r-1)\cK}\left(e^{n\cK} -P\right)x\right\| \le  \left\|(e^{n\cK} -P)x\right\| \xrightarrow[n\to\infty]{}0, 
\end{align*}
where we have used $e^{n(r-1)\cK}P = P$ and the fact that $\cK$ generates a contraction semigroup. Hence, $M_r$ and $\cL_{n,r}$ satisfy the assumptions of
Theorem~\ref{thm:StrongBinoFormula} and by applying this to the expansion \eqref{eq:Expansion} we see that
\begin{align}
\label{eq:Established}
\lim_{n\to\infty}e^{nr\cK + \cL}x = e^{P\cL_rP}Px = e^{P\cL P}Px,
\end{align}
with convergence uniformly in $r\in[1,2]$,
where we have used the following:
\begin{align*}
    P\cL_rP = \int_0^1 Pe^{sr\cK}\cL e^{(1-s)r\cK}Pds = P\cL P.
\end{align*}
To conclude the proof of \eqref{eq:StrongDampingConv} let now $\eps>0$. By \eqref{eq:Established} we know that there exists some $N\in\N$ such that for all $n\ge N$
\begin{align*}
\sup_{r\in[1,2]}\left\|\left(e^{nr\cK + \cL} -  e^{P\cL P}P\right)x\right\| \le \eps.
\end{align*}
Let now  $\gamma\ge N$ be a real number and note that for $n_\gamma=\floor{\gamma}\ge N$ we have $r_\gamma =\gamma /n_\gamma \in[1,2]$. This gives
\begin{align*}
\left\|\left(e^{\gamma\cK + \cL} -  e^{P\cL P}P\right)x\right\| = \left\|\left(e^{n_\gamma r_\gamma \cK + \cL} -  e^{P\cL P}P\right)x\right\| \le  \eps.
\end{align*}
Since $\eps>0$ was arbitrary, this finishes the proof of \eqref{eq:StrongDampingConv}. 

We complete the proof of the theorem by showing the bound on the speed of convergence \eqref{eq:StrongDampSpeed}. For that we use \eqref{eq:SpeedBino}  of Theorem~\ref{thm:StrongBinoFormula} together with the fact that by \eqref{eq:LSpeedDamp} we have $s_n'\le c/n$ for some $c\ge 0$, which gives for all $\mathbf{N}=\left(N_l\right)_{l\in\N} \subseteq\N$ and some $C\ge 1$
\begin{align}
\label{eq:KgammaBound}
\nn\left\|\left(e^{\gamma\cK + \cL} -  e^{P\cL P}P\right)x\right\| &= \left\|\left(e^{n_\gamma r_\gamma \cK + \cL} -  e^{P\cL P}P\right)x\right\| = \left\|\left(M_{r_\gamma} + \frac{\cL_{n_\gamma,r_\gamma}}{n_\gamma}\right)^{n_\gamma}x\right\|\\& \le C \sup_{l\in\N}\left(\left(\sup_{\tilde r\in [0,2]}\|\cL_{\tilde r}\|\right)^{-l+1}s_{N_l}\left((\cL_{r_\gamma}P)^{l-1}x\right) +\frac{N_l}{n_\gamma}\|x\|\right),
\end{align}
where we have denoted the speed of convergence for the operators $M_{r} = e^{r
\cK}$ as defined in \eqref{eq:Mpowerspeed} by
\begin{align*}
    s_n(x) = \sup_{n'\ge n,r \in [1,2]}\left\|\left(M^{n'}_r-P\right)x\right\|.
\end{align*}
Note that 
\begin{align*}
    s_{N_l}\left((\cL_{r_\gamma}P)^{l-1}x\right) &= \sup_{n'\ge N_l,r\in [1,2]}\left\|\left(M^{n'}_{r}-P\right)\int_0^1 e^{sr_\gamma\cK}\cL\,ds\,(P\cL P)^{l-2}x\right\| \\&= \sup_{n'\ge N_l,r\in [1,2]}\left\|\int_0^1\left(e^{(n'r+sr_\gamma )\cK}-P\right)\,ds(\cL P)^{l-1}x\right\| \\&\le \sup_{\gamma'\ge N_l,}\left\|\left(e^{\gamma'\cK}-P\right)(\cL P)^{l-1}x\right\| = v_{N_l}\left((\cL P)^{l-1}x\right),
\end{align*}
with $v_\gamma(x)$ being defined in \eqref{eq:Kspeed}.
Moreover, since $\cK$ generates a contraction semigroup we have
\begin{align*}
    \left\|\cL_r\right\| = \left\|\int_0^1 e^{sr\cK}\cL e^{(1-s)r\cK}ds\right\| \le \left\|\cL\right\|
\end{align*}
and hence $\sup_{r\in[0,2]}\left\|\cL_r\right\| = \left\|\cL\right\|. $ %Plugging both 
Substituting these in \eqref{eq:KgammaBound}, we obtain 
\begin{align*}
\left\|\left(e^{\gamma\cK + \cL} -  e^{P\cL P}P\right)x\right\| \le C \sup_{l\in\N}\left(\|\cL\|^{-l+1}v_{N_l}\left((\cL P)^{l-1}x\right) +\frac{N_l}{n_\gamma}\|x\|\right).
\end{align*}

\end{proof}
\section{Discussion of speed of convergence in Zeno and strong damping limit}
\label{sec:ExamplesZeno}
In this section we discuss cases and examples of contractions $M$ and contraction semigroups $\left(e^{t\cK}\right)_{t\ge 0}$ for which the provided bounds on the speed of convergence, $s^{\sup}_n(x,t\cL,\mathbf{N})$ in \eqref{eq:SpeedZeno} of Theorem~\ref{thm:StrongZenoGeneralQuant} and $v^{\sup}_\gamma(x,t\cL,\mathbf{N})$ in \eqref{eq:StrongDampSpeed} of Theorem~\ref{thm:StrongDamping}, can be controlled explicitly. As outlined in Remark~\ref{rem:SpeedOfConv}, the idea is to have good control on the speed of the strong power convergence $M^nx\xrightarrow[n\to\infty]{}Px$ and the strong mixing $e^{\gamma\cK}x\xrightarrow[\gamma\to\infty]{}Px$, i.e.~using the notation of \eqref{eq:MpowerspeedZeno} and \eqref{eq:Kspeed}, on 
 \begin{align}
 \label{eq:MixingSpeedi}
   s_n(x) =\sup_{n'\ge n}\|(M^{n'}-P)x\|\quad\quad\text{and}\quad\quad v_{\gamma}(x) = \sup_{\gamma'\ge\gamma}\left\|\left(e^{\gamma' \cK} -P\right)x\right\|,
 \end{align}
 respectively. This can then be utilised to pick a reasonable $n$-dependent $\mathbf{N}(n)$ (or $\gamma$ dependent $\mathbf{N}(\gamma)$) ensuring that also the convergence speeds
 \begin{align*}
     s^{\sup}_n(x,t\cL,\mathbf{N}(n)) \quad\quad\text{and}\quad\quad v^{\sup}_\gamma(x,t\cL,\mathbf{N}(\gamma))
 \end{align*}
 of the Zeno- and strong damping limits decay nicely as $n\to\infy$ and $\gamma\to\infy$, respectively.

Firstly, we consider the case of uniformly power convergent $M$ and uniformly mixing $\left(e^{t\cK}\right)_{t\ge0}$ , i.e.~for which $M^n\xrightarrow[n\to\infty]{}P$ and $e^{\gamma\cK}\xrightarrow[\gamma\to\infty]{}P$, does not only hold in strong topology but even in uniform topology. After that, we consider the so-called \emph{(bosonic quantum limited) attenuator channel and semigroup.} In both cases we find that the sequences in \eqref{eq:MixingSpeedi} decay exponentially fast. This then implies bounds on the speed of convergence for the respective Zeno- and strong damping limits 
of order $\mathcal{O}(\log(n)/n)$ and $\mathcal{O}(\log(\gamma)/\gamma)$, respectively.

\subsection{Uniform Zeno- and strong damping limits}

Here, we consider the case in which the convergences
\begin{align}
\label{eq:MixingUniformSec}
M^n\xrightarrow[n\to\infty]{}P\quad\quad \text{and}\quad\quad e^{\gamma\cK}\xrightarrow[\gamma\to\infty]{}P
\end{align} 
do not only hold in strong- but even in uniform topology. In that case, using \cite[Corollary 3.2]{BeckerDattaSalz_Zeno_2021} and  \cite[Corollary 4.1.2]{Salzmann_PhDthesis_2023} respectively, we see that the convergences in \eqref{eq:MixingUniformSec} happen in fact even exponentially fast. In the language of Section~\ref{sec:ZenoAndStrongFromBino} using the notation \eqref{eq:SpeedZeno} and \eqref{eq:Kspeed} established there, this would give for the abstract speeds of convergence  
\begin{align}
    \sup_{\|x\|=1} s_n(x) \le C\delta^n\quad\quad\text{and}\quad\quad\sup_{\|x\|=1} v_\gamma(x) \le C e^{-c\gamma}
\end{align}
for some $0< \delta<1$ and $C,c>0.$ 
In the proof of Theorem~\ref{thm:UniformZenoGeneralQuant} and~\ref{thm:UniformStrongDamping} below we combine this with Theorem~\ref{thm:StrongZenoGeneralQuant} and~\ref{thm:StrongDamping} respectively to show bounds on the speed of convergence of the respective Zeno- and strong damping limits.

\begin{theorem}[Zeno limit in uniform topology]
\label{thm:UniformZenoGeneralQuant}
Let $M\in\cB(X)$ be a contraction which satisfies in operator norm
\begin{align}
\label{eq:UniformPowerConvZenothm}
\lim_{n\to\infty}M^n  =P 
\end{align}
for some operator $P\in\cB(X)$. Then for $\cL\in\cB(X)$ we have
\begin{align}
\lim_{n\to\infty}\left(Me^{t\cL/n}\right)^n  = e^{tP\cL P}P.
\end{align}
 More precisely, we have for all $n\ge 2$
\begin{equation}
\label{eq:SpeedZenoUniform}
    \left\|\left(Me^{t\cL/n}\right)^n - e^{tP\cL P}P\right\|\le \frac{C\log(n)}{n},
\end{equation}
for some $C\ge 1$ independent of $n.$
\end{theorem}
\begin{remark}
In \cite{möbus2022optimal} it was shown that for uniformly power convergent contractions $M,$ the Zeno product converges with convergence speed $\mathcal{O}(1/n).$ Moreover, the authors have proven that this is the optimal convergence speed. To contrast this, note that
Theorem~\ref{thm:UniformZenoGeneralQuant} shows that using our bound on the speed of convergence of the Zeno product provided in Theorem~\ref{thm:StrongZenoGeneralQuant} in this context gives a bound on the convergence speed of order $\mathcal{O}(\log(n)/n)$ for the corresponding Zeno limit. Hence, the bound of Theorem~\ref{thm:UniformZenoGeneralQuant} is optimal up to logarithmic factors in $n.$
\end{remark}
\begin{proof}[Proof of Theorem~\ref{thm:UniformZenoGeneralQuant}]
It suffices to prove \eqref{eq:SpeedZenoUniform}. Using \cite[Corollary 3.2]{BeckerDattaSalz_Zeno_2021} we know that there exists $c\ge 0$ and $0<\delta<1$ such that for all $n\in\N$
\begin{align*}
\left\|M^n-P\right\| \le c\, \delta ^n. 
\end{align*}
Hence, using the notation established in \eqref{eq:MpowerspeedZeno} we see that for all $x\in X$ we have
\begin{align*}
s_n(x) \le c\,\delta^n\|x\|.
\end{align*}
Therefore, using \eqref{eq:SpeedZeno} in Theorem~\ref{thm:StrongZenoGeneralQuant} we see for all $(N_l)_{l\in\N}\subseteq \N$ and some $C\ge 1$ 
\begin{align*}
\left\|\left(\left(Me^{t\cL/n}\right)^n - e^{tP\cL P}P \right)x\right\| &\le C\sup_{l\in\N}\left(\|t\cL\|^{-l+1}s_{N_l}\left((t\cL P)^{l-1}x\right) +\frac{N_l}{n}\|x\|\right) \\&\le C\sup_{l\in\N}\left(c\delta^{N_l} + \frac{N_l}{n}\right)\|x\|,
\end{align*}
where we have used that $P$ is a contraction. Picking now the constant sequence 
\begin{align*}
N_l= \floor{\frac{\log(n)}{\log\left(1/\delta\right)}}
\end{align*}
for all $l\in\N,$ we see
\begin{align*}
\left\|\left(Me^{t\cL/n}\right)^n - e^{tP\cL P}P \right\| \le C \left(c\delta^{\frac{\log(n)}{\log\left(1/\delta\right)}-1} + \frac{\log(n)}{\log\left(1/\delta\right)n}\right) = C \left(\frac{c}{n\delta} + \frac{\log(n)}{\log\left(1/\delta\right)n}\right).
\end{align*}
Renaming the constant $C$ yields \eqref{eq:SpeedZenoUniform}.
\end{proof}
\begin{theorem}[Strong damping in uniform topology]
	\label{thm:UniformStrongDamping}
	Let $\cK$ with domain $\cD(\cK)\subseteq X$ be the generator of a strongly continuous contraction semigroup which satisfies in operator norm
	\begin{align}
	\label{eq:StrongErgodicityUni}
	\lim_{\gamma\to\infty} e^{\gamma \cK} = P,
	\end{align}
	for some $P\in\cB(X).$ Furthermore, let $\cL\in\cB(X)$. Then for all $t >0$
	\begin{align}
	\label{eq:StrongDampingConvUni}
	\lim_{\gamma\to\infty}e^{t(\gamma \cK+\cL)} = e^{tP\cL P}P.
	\end{align}
 More precisely, we have for all $\gamma\ge 2$
 \begin{align}
 \label{eq:StrongDampSpeedUniform}
\left\|e^{t(\gamma\cK + \cL)} -  e^{tP\cL P}P\right\| \le \frac{C\log(\gamma)}{\gamma},
\end{align}
for some $C\ge 1$ independent of $\gamma.$
\end{theorem}
\begin{proof}[Proof of Theorem~\ref{thm:UniformStrongDamping}]
The proof follows exactly the same lines as the one of Theorem~\ref{thm:UniformZenoGeneralQuant}. But for completeness we give the details here explicitly.

It suffices to prove \eqref{eq:StrongDampSpeedUniform}. Using \cite[Corollory 4.1.2]{Salzmann_PhDthesis_2023} we know that there exists $c_1,c_2> 0$ such that for all $\gamma\ge0$
\begin{align*}
\left\|e^{\gamma\cK}-P\right\| \le c_1\, e^{-c_2 \gamma}. 
\end{align*}
Hence, using the notation established in \eqref{eq:Kspeed} we see that for all $x\in X$ we have
\begin{align*}
v_\gamma(x) \le c_1\,e^{-c_2 \gamma}\|x\|.
\end{align*}
Therefore, using \eqref{eq:StrongDampSpeed} in Theorem~\ref{thm:StrongDamping} we see for all $(N_l)_{l\in\N}\subseteq \N$ and some $C\ge 1$ 
\begin{align*}
\left\|\left(e^{t(\gamma\cK + \cL)} -  e^{tP\cL P}P \right)x\right\| &\le C\sup_{l\in\N}\left(\|t\cL\|^{-l+1}v_{N_l}\left((t\cL P)^{l-1}x\right) +\frac{N_l}{\floor{\gamma}}\|x\|\right)\\&\le C\sup_{l\in\N}\left(c_1e^{-c_2N_l} + \frac{N_l}{\floor{\gamma}}\right)\|x\|,
\end{align*}
where we have used that $P$ is a contraction. Picking now for $\gamma\ge 2$ the constant sequence 
\begin{align*}
N_l= \floor{\frac{\log(\gamma)}{c_2}}
\end{align*}
for all $l\in\N,$ we see
\begin{align*}
\left\|e^{t(\gamma\cK + \cL)} -  e^{tP\cL P}P \right\| \le C \left(c_1e^{-\log(\gamma)+c_2} + \frac{\log(\gamma)}{c_2 \floor{\gamma}}\right) = C \left(\frac{c_1e^{c_2}}{\gamma} + \frac{\log(\gamma)}{c_2 \floor{\gamma}}\right).
\end{align*}
Using $\frac{1}{\floor{\gamma}}\le \frac{2}{\gamma}$ and renaming the constant $C$ yields \eqref{eq:StrongDampSpeedUniform}.
\end{proof}
\subsection{Bosonic quantum limited attenuator channel}
\label{sec:Atten}
In the following we discuss the example of the so-called \emph{(bosonic quantum limited) attenuator channel} and the corresponding \emph{attenuator semigroup}, which can be seen to model a beam splitter, as highlighted below. Notably, the considered channel and semigroup will respectively be strongly-, but not uniformly power convergent and mixing. Using our Theorem~\ref{thm:StrongZenoGeneralQuant} and Theorem~\ref{thm:StrongDamping} we can establish Zeno- and strong damping limits for the attenuator. Moreover, as we see in Lemma~\ref{lem:AttenMixing} and the discussion around \eqref{eq:AttStrongPowerConv1}, the attenuator is strongly power convergent and mixing with exponential convergence speed, provided that the input state has finite particle number (see \eqref{eq:AverageParticleNumber} for definition). Similarly to the discussion in the last section, this can then be combined with the bounds on the speed of convergence provided in Theorem~\ref{thm:StrongZenoGeneralQuant} and~\ref{thm:StrongDamping} to show convergence speed of the respective Zeno- and strong damping limits of order $\mathcal{O}(\log(n)/n)$ and $\mathcal{O}(\log(\gamma)/\gamma)$ respectively (see Proposition~\ref{prop:AttZenoDamp}).

We first give some basic definitions and facts corresponding to the attenuator channel in Lemma~\ref{lem:AttBasics} and show its continuity properties in Lemma~\ref{lem:AttenMixing}. The latter will be essential to show the exponential strong power convergence and mixing mentioned above. Then we give the main result of this section in Proposition~\ref{prop:AttZenoDamp}.

\bigskip

For that, let $\cH$ be some infinite-dimensional, separable Hilbert space with orthonormal basis $\left(\ket{n}\right)_{n\in\N_0}.$ We define the number operator $N:\cD(N) \to \cH$ by its action on the basis vectors $\ket{n}$ as 
\begin{align*}
N\ket{n} = n\ket{n}    
\end{align*}
and then extended linearly. Here, we have denoted the dense domain $\cD(N)\subseteq\cH$ which is defined as
\begin{align*}
    \cD(N) = \left\{\psi\in\cH\Big|\, \sum_{n=0}^\infty n^2|\bra{n}\psi\rangle|^2 <\infty\right\}. 
\end{align*}
For a state $\rho$ (or more generally $\rho\in\cT(\cH)_+$)  we define its \emph{(average) particle number}, by \cite{BeckerDatta_ConvRatesEnergDiamon_2019,Shirokov_energy-constrained_2018}
\begin{align}
\label{eq:AverageParticleNumber}
    \Tr\left(N\rho\right) \coloneqq \sum_{n=0}^\infy n\bra{n}\rho\ket{n}.
\end{align}
Note that the series as a sum of non-negative numbers is well-defined in $[0,\infty]$.\footnote{Note that $N$ is unbounded and so $N\rho$ in general not well-defined and especially not trace class. Therefore, the expression $\Tr(N\rho)$ needs to be understood in the formal sense. However, we can make sense of it by formally evaluating the trace in the eigenbasis of the number operator $\left(\ket{n}\right)_{n\in\N}$ as on the right hand side of \eqref{eq:AverageParticleNumber}.} 

For each $\alpha\in\C$ we can define the \emph{coherent state (vector)}, see e.g.~\cite[Chapter 3]{Gerryknight_QuantumOptics_2004} or \cite[Chapter 3.10]{Holevo_ProbabilisticQuantumBook_2011},
\begin{align*}
    \ket{\alpha} = e^{-\frac{|\alpha|^2}{2}}\sum^{\infty}_{n=0} \frac{\alpha^n}{\sqrt{n!}}\ket{n}.
\end{align*}  
The coherent states satisfy the elementary relations
\begin{align}
\label{eq:CoherentOverlap}
\bra{\alpha}\beta\rangle = e^{-(|\alpha|^2+|\beta|^2 - 2\alpha^*\beta)}
\end{align}
for all $\alpha,\beta\in\C$ and 
\begin{align}
\label{eq:CoherenIdent}
\frac{1}{\pi}\int_\C\kb{\alpha}d\alpha = \1, 
\end{align}
where $d\alpha=d\Re(\alpha)d\Im(\alpha),$ which can be checked by direct calculation. Note that due to \eqref{eq:CoherentOverlap} and \eqref{eq:CoherenIdent} the coherent states can be interpreted as an \emph{overcomplete basis.}

For $\eta\in\C$ with $|\eta|\le1$, we can define the \emph{(single mode bosonic quantum limited) attenuator channel} \cite{DePalmaGiovanneti_Attenuator_2016,Winter_EnergyConstrDiamond_2017} by its action on coherent states as 
\begin{align}
\label{eq:CoherentAtt}
    \Phi^{\text{att}}_\eta(\kb{\alpha}) = \kb{\eta\alpha}.
\end{align} 
The Lemma~\ref{lem:AttBasics} below shows some  basic statements about the attenuator channel which in some form have also been stated in \cite{DePalmaGiovanneti_Attenuator_2016,Winter_EnergyConstrDiamond_2017,BeckerDatta_ConvRatesEnergDiamon_2019}. In particular we show that the attenuator channel is already uniquely determined by the above relation \eqref{eq:CoherentAtt}, which essentially follows using that the coherent states form an overcomplete basis. Moreover, from the %relation of 
expression for its Stinespring isometry~\eqref{eq:AttStinespring} it becomes apparent that the attenuator channel can be considered to model a beam splitter with transmitivity $\eta,$ where one input beam is given by the channel input and the other is given by the \emph{vacuum state} $\kb{0}$ (compare \cite[Chapter 6.2]{Gerryknight_QuantumOptics_2004} and particularly Equation~(6.16) therein). 
\begin{lemma}
\label{lem:AttBasics}
For $\eta\in\C$ with $|\eta|\le1$ the relation \eqref{eq:CoherentAtt} uniquely defines a quantum channel $\Phi^{\text{att}}_\eta:\cT(\cH) \to \cT(\cH).$ Its Stinespring isometry $V_\eta: \cH\to\cH\otimes\cH_E$ is, up to isometrically changing the enviroment Hilbert space $\cH_E$, determined by its action on coherent states as
\begin{align}
\label{eq:AttStinespring}
    V_\eta\ket{\alpha} = \big|\eta\alpha\big\rangle\big|\sqrt{1-|\eta|^2}\,\alpha\big\rangle.
\end{align}
Moreover, the attenuator channel satisfies the semigroup property, i.e.~for all $\eta_1,\eta_2\in \C$ with $|\eta_1|,|\eta_2|\le1$ we have
\begin{align}
\label{eq:AttSemigroup}
\Phi^{\text{att}}_{\eta_1\eta_2} = \Phi^{\text{att}}_{\eta_1}\circ\Phi^{\text{att}}_{\eta_2}.
\end{align}
Lastly, $\Phi^{\text{att}}_\eta$ can be written in Kraus decomposition as 
\begin{align}
\label{eq:AttKraus}
   \Phi^{\text{att}}_\eta(x) =\sum_{l=0}^\infy K_l(\eta)xK^*_l(\eta),
\end{align}
where $K_l(\eta) = \sum_{n=0}^\infty \sqrt{{n+l \choose n}\left(1-|\eta|^2\right)^{l} }\,\eta^n \ket{n}\!\bra{n+l}.$
\end{lemma}
\begin{proof}[Proof of Lemma~\ref{lem:AttBasics} ]
We start by proving existence and uniquenes of the quantum channel $\Phi^{\text{att}}_\eta.$ For existence note that \eqref{eq:AttStinespring} defines, after extending by linearity using the relation \eqref{eq:CoherenIdent}, a linear map 
$V_\eta:\cH \to \cH\otimes\cH_E$ (where $\cH_E \cong \cH$) explicitly given for $\psi\in\cH$ by \begin{align}
    \label{eq:StinespringAttFull}
    V_\eta\psi = \frac{1}{\pi}\int V_\eta\kb{\alpha}\psi\rangle d\alpha= \frac{1}{\pi}\int \ket{\eta\alpha}\ket{\sqrt{1-|\eta|^2}\,\alpha}\bra{\alpha}\psi\rangle d\alpha.
\end{align}
Moreover, $V_\eta$ is in fact an isometry as
\begin{align*}
    \left\|V_\eta\psi\right\|^2 &=
\frac{1}{\pi^2}\iint \bra{\beta}\psi\rangle\!\bra{\psi}\alpha\rangle \bra{\eta\alpha}\eta\beta\rangle\bra{\sqrt{1-|\eta|^2}\,\alpha}\sqrt{1-|\eta|^2}\,\beta\rangle\,d\alpha d\beta \\&= \frac{1}{\pi^2}\iint \bra{\beta}\psi\rangle\!\bra{\psi}\alpha\rangle e^{-|\eta|^2(|\alpha|^2+|\beta|^2 -2\alpha^*\beta)}e^{-(1-|\eta|^2)(|\alpha|^2+|\beta|^2 -2\alpha^*\beta)} \,d\alpha d\beta \\&=\frac{1}{\pi^2}\iint \bra{\beta}\psi\rangle\!\bra{\psi}\alpha\rangle \bra{\alpha}\beta\rangle \,d\alpha d\beta = \|\psi\|^2,
\end{align*}
where we have used the relation \eqref{eq:CoherentOverlap} twice. Hence, $\Phi^{\text{att}}_\eta(\placeholder) = \Tr_E\left(V_\eta(\placeholder) V^*_\eta\right)$ defines a quantum channel, which by definition satisfies \eqref{eq:CoherentAtt}.

For uniqueness note that any channel $\Phi'_{\eta}$ satisfying \eqref{eq:CoherentAtt}, in particular mapping a set of pure states onto pure states, must have a Stinespring isometry $V'_\eta:\cH\to\cH\otimes\cH_E$ such that 
$
    V'_\eta\ket{\alpha} = \ket{\eta\alpha}\ket{\phi_{\alpha,\eta}}
$
for some state vector $\ket{\phi_{\alpha,\eta}} \in \cH_E.$ By isometry, we already know for $\alpha,\beta\in\C$ that
$\bra{\beta}\alpha\rangle = \bra{\eta\beta}\eta\alpha\rangle\bra{\phi_{\beta,\eta}}\phi_{\alpha,\eta}\rangle$ and hence 
\begin{align*}
\bra{\phi_{\alpha,\eta}}\phi_{\beta,\eta}\rangle = \bra{\alpha}\beta\rangle/\bra{\eta\alpha}\eta\beta\rangle = e^{-(1-\eta^2)(|\alpha|^2+|\beta|^2 - 2\alpha^*\beta)} = \bra{\sqrt{1-|\eta|^2}\,\alpha}\sqrt{1-|\eta|^2}\,\beta\rangle.
\end{align*}
From that we see that $V_\eta$ and $V'_\eta$ are equal (up to isometrically changing the environment Hilbert space) and in particular the actions of the channels are equal, i.e.~for $x\in \cT(\cH)$ we have
\begin{align*}
\Phi'_\eta(x) &= \Tr_E(V'_\eta x{V'}_\eta^*)= 
\frac{1}{\pi^2}\iint\bra{\beta}x\ket{\alpha} \Tr_E(V'_\eta\ket{\beta}\!\bra{\alpha}{V'}_\eta^*)\,d\alpha d\beta\\&= \frac{1}{\pi^2}\iint\bra{\beta}x\ket{\alpha} \ket{\eta\beta}\!\bra{\eta\alpha}\bra{\phi_{\alpha,\eta}}\phi_{\beta,\eta}\rangle d\alpha d\beta =\Phi^{\text{att}}_\eta(x).
\end{align*}
We proceed to prove the semigroup property \eqref{eq:AttSemigroup}. For that note that 
\begin{align*}
    \left(V_{\eta_2}\otimes\1_E \right)V_{\eta_1}\ket{\alpha} = \left(V_{\eta_2}\ket{\eta_1\alpha}\right)\ket{\sqrt{1-|\eta_1|^2}\,\alpha} = \ket{\eta_1\eta_2\alpha}\ket{\sqrt{1-|\eta_1|^2}\,\alpha}\ket{\sqrt{1-|\eta_2|^2}\,\eta_1\alpha},
\end{align*}
where we denoted the identity on the environment Hilbert space $\cH_E$ by $\1_E.$
The semigroup property then follows using the same argument as above together with the fact that 
\begin{align*}
    \bra{\sqrt{1-|\eta_1|^2}\,\alpha}\sqrt{1-|\eta_1|^2}\,\beta\rangle\bra{\sqrt{1-|\eta_2|^2}\,\eta_1\alpha}\sqrt{1-|\eta_2|^2}\,\eta_1\beta\rangle = \bra{\sqrt{1-|\eta_1\eta_2|^2}\,\alpha}\sqrt{1-|\eta_1\eta_2|^2}\,\beta\rangle,
\end{align*}
which directly follows from \eqref{eq:CoherentOverlap}.

Lastly we prove \eqref{eq:AttKraus}. First we convince ourselves that the right hand side of \eqref{eq:AttKraus} indeed defines a quantum channel. Here, complete positivity is evident as the map is already written in Kraus decomposition. The trace preservation property follows from
\begin{align*}
    \sum_{l=0}^\infty K_l^*(\eta)K_l(\eta) &= \sum_{l=0}^\infty\sum_{n=0}^\infty{n+l \choose n} \left(1-|\eta|^2\right)^{l} |\eta|^{2n} \kb{n+l}\\&= \sum_{m=0}^{\infty}\sum_{k=0}^m {m \choose k} \left(1-|\eta|^2\right)^{m-k} |\eta|^{2k}\kb{m} = \1.
\end{align*}
For the equality of the two channels in \eqref{eq:AttKraus} it suffices to show equality 
on coherent states by the established uniqueness. For that see
\begin{align*}
&\sum_{l=0}^\infty K_l(\eta)\kb{\alpha}K^*_l(\eta)\\&=  e^{-|\alpha|^2}\sum_{l=0}^\infty\sum_{m=0}^\infty\sum_{n=0}^\infty\sqrt{{n+l \choose n} {m+l \choose m} }(1-|\eta|^2)^l \,{\eta^*}^{n}\eta^{m}\frac{{\alpha^*}^n{\alpha}^m|\alpha|^{2l}}{\sqrt{(n+l)!(m+l)!}}\,\ket{m}\!\bra{n} \\&= e^{-|\eta\alpha|^2} \sum_{m=0}^\infty\sum_{n=0}^\infty \,{\eta^*}^{n}\eta^{m}\frac{{\alpha^*}^n{\alpha}^m}{\sqrt{n!m!}}\,\ket{m}\!\bra{n} = \kb{\eta\alpha}.
\end{align*}
Hence, the quantum channel defined by the right hand side of \eqref{eq:AttKraus} satisfies \eqref{eq:CoherentAtt} and is therefore equal to $\Phi^{\text{att}}_\eta$ on all of $\cT(\cH).$ 

\end{proof}

As shown in \cite[Proposition 1]{Winter_EnergyConstrDiamond_2017} the map $\eta\mapsto\Phi^{\text{att}}_\eta$ is not uniformly continuous. In fact for all $\eta_1\neq\eta_2$ we have
\begin{align}
\label{eq:AttNotUniformCont}
    \left\|\Phi^{\text{att}}_{\eta_1} - \Phi^{\text{att}}_{\eta_2}\right\| = 2.
\end{align}
This can easily be seen by lower bounding the operator norm by the corresponding supremum over all coherent states as
\begin{align*}
    \left\|\Phi^{\text{att}}_{\eta_1} - \Phi^{\text{att}}_{\eta_2}\right\| &= \sup_{\|x\|_1=1}\left\|\left(\Phi^{\text{att}}_{\eta_1} - \Phi^{\text{att}}_{\eta_2}\right)(x)\right\|_1 \ge \sup_{\alpha\in\C}\left\|\left(\Phi^{\text{att}}_{\eta_1} - \Phi^{\text{att}}_{\eta_2}\right)(\kb{\alpha})\right\|_1 \\&= 2\sup_{\alpha\in\C}\sqrt{1-|\langle \eta_1\alpha|\eta_2 \alpha\rangle|^2} = 2,
\end{align*}
where for the second to last equation we used the expression of the trace distance for pure states \eqref{eq:PureTraceDistance} and the last equation follows by \eqref{eq:CoherentOverlap}.

However, the map $\eta\mapsto\Phi^{\text{att}}_\eta$ can be shown to be strongly continuous. Moreover, for input states which have a finite particle number we even get the explicit continuity bound \eqref{eq:AttMixingBound} for continuity at $\eta=0.$ This is the content of the following lemma.

\begin{lemma}
	\label{lem:AttenMixing}
 The map $\eta\mapsto \Phi^{\text{att}}_\eta$ is strongly continuous, i.e.~for all $x\in\cT(\cH)$ we have continuity of the map
 \begin{align}
     \eta\mapsto \Phi^{\text{att}}_\eta(x).
 \end{align}
 Moreover, we have for $\rho\in\cT(\cH)_+$  the explicit continuity bound
\begin{equation}
\label{eq:AttMixingBound}
	\left\|\Phi^{\text{att}}_\eta(\rho) - P(\rho)\right\|_1 \le 4|\eta|\, \Tr\left((N+\1)\rho\right).
\end{equation}
where we have denoted $P(\rho)= \Phi^{\text{att}}_0(\rho) = \Tr(\rho)\,\kb{0}.$

\end{lemma}
\begin{proof}
We start by proving the strong continuity of $\eta\mapsto\Phi^{\text{att}}_\eta.$ For that note first that the map $\alpha \mapsto \ket{\alpha}$ is continuous which can be seen by using \eqref{eq:CoherentOverlap} and therefore
\begin{align*}
    \left\|\ket{\alpha}-\ket{\beta}\right\|^2 = 2\left(1-\Re\left(\bra{\alpha}\beta\rangle\right)\right) = 2\left(1- e^{-(|\alpha|^2+|\beta|^2 - 2\alpha^*\beta)}\right),
\end{align*}
 and by that $\lim_{\alpha\to\beta}\ket{\alpha}=\ket{\beta}.$ By that we also get for all $\alpha,\beta\in\C$ continuity of the map
\begin{align}
\label{eq:Weirdalphabetacont}
    \nn B_1(0) &\to \cT(\cH)\\
    \eta &\mapsto \ket{\eta\beta}\!\bra{\eta\alpha}\bra{\sqrt{1-|\eta|^2}\,\alpha}\sqrt{1-|\eta|^2}\,\beta\rangle,
\end{align}
where we have denoted the unit ball in the complex numbers by $B_1(0) = \{\eta\in\C | \,|\eta|\le1\}$.
Now note that for any $n,m\in\N_0$ we have
\begin{align*}
    \iint |\bra{\beta} n\rangle\!\langle m\ket{\alpha}|\,d\alpha d\beta =  \iint e^{-(|\alpha|^2+|\beta|^2)/2}\frac{|\beta|^n|\alpha|^m}{\sqrt{n! m!}}\,d\alpha d\beta <\infty.
\end{align*}
In particular, this gives by the triangle inequality that for all $x\in\text{span}\left\{\ket{n}\!\bra{m}\ \big|\,n,m\in\N_0\right\}$ the function $(\alpha,\beta) \mapsto |\bra{\beta}x\ket{\alpha}|$ is integrable. Therefore, using continuity of the map \eqref{eq:Weirdalphabetacont}  together with the dominated convergence theorem gives continuity of
\begin{align*}
    \eta \mapsto \Phi^{\text{att}}_\eta(x) = \frac{1}{\pi^2}\iint\bra{\beta}x\ket{\alpha} \ket{\eta\beta}\!\bra{\eta\alpha} \bra{\sqrt{1-|\eta|^2}\,\alpha}\sqrt{1-|\eta|^2}\,\beta\rangle\, d\alpha d\beta 
\end{align*}
for all $x\in\text{span}\left\{\ket{n}\!\bra{m}\,|\,n,m\in\N_0\right\}.$ However, since $\text{span}\left\{\ket{n}\!\bra{m}\,|\,n,m\in\N_0\right\}\subseteq\cT(\cH)$ is dense and $\Phi^{\text{att}}_\eta$ is a contraction for all $\eta\in B_1(0)$, this already establishes continuity of $\eta \mapsto \Phi^{\text{att}}_\eta(x)$ for all $x\in\cT(\cH).$
\\\indent We know proceed to prove the explicit continuity bound \eqref{eq:AttMixingBound}. It suffices to show the statement for $\rho$ being a state as it then immediately follows for general $\rho\in\cT(\cH)_+$ by homogeneity of left and right hand side of \eqref{eq:AttMixingBound}.  Hence, let $\rho$ be a state and write by the triangle inequality
	\begin{align}
		\label{eq:AttTriangle}
		\nn	\left\|\Phi^{\text{att}}_\eta(\rho) - P(\rho)\right\|_1 &\le \left|\bra{0}\left(\Phi^{\text{att}}_\eta(\rho) - P(\rho)\right)\ket{0}\right| + 	\left\|(\1-\kb{0})\Phi^{\text{att}}_\eta(\rho)(\1-\kb{0})\right\|_1 \\&+\left\|\kb{0}\Phi^{\text{att}}_\eta(\rho)(\1-\kb{0})\right\|_1+ \left\|(\1-\kb{0})\Phi^{\text{att}}_\eta(\rho) \kb{0}\right\|_1 .
	\end{align}
	For the first term we use the Kraus decomposition \eqref{eq:AttKraus} which gives
	\begin{align}
		\label{eq:Att0Term}
		\nn\left|\bra{0}\Phi^{\text{att}}_\eta(\rho) - P(\rho)\ket{0}\right| &=  1 - \bra{0}\Phi^{\text{att}}_\eta(\rho)\ket{0} = \sum_{l=0}^\infty \left(1-\left(1-|\eta|^2\right)^l\right)\bra{l}\rho\ket{l} \\&\le |\eta|^2\sum_{l=0}^\infty l \bra{l}\rho\ket{l}  = |\eta|^2\Tr(N\rho),
	\end{align}
	where for the inequality we have used
	\begin{align*}
	1-\left(1-|\eta|^2\right)^l \le l |\eta|^2,
	\end{align*}
 which follows by the mean value theorem.
	For the second term in \eqref{eq:AttTriangle} we have by positivity of $\Phi^{\text{att}}_\eta(\rho)$ that
	\begin{align}
		\label{eq:Att1-0}
	\nn\Big\|(\1-\kb{0})\Phi^{\text{att}}_\eta(\rho)(\1-\kb{0})\Big\|_1 &= \Tr\Big((\1-\kb{0})\Phi^{\text{att}}_\eta(\rho)\Big) = 1- \bra{0}\Phi^{\text{att}}_\eta(\rho)\ket{0}\\& \le |\eta|^2\Tr(N\rho),
	\end{align}
	where we have used that $\Phi^{\text{att}}_\eta$ is trace preserving and for the last inequality the estimate in \eqref{eq:Att0Term}.
For the cross terms we then use H\"older's inequality and \eqref{eq:Att1-0} to see
\begin{align*}
	\left\|\kb{0}\Phi^{\text{att}}_\eta(\rho)(\1-\kb{0})\right\|_1 & \le \left\|\kb{0}\Phi^{\text{att}}_\eta(\rho)^{1/2}\right\|_2\left\|\Phi^{\text{att}}_\eta(\rho)^{1/2}(\1-\kb{0})\right\|_2 \\& = \sqrt{\left\|\kb{0}\Phi^{\text{att}}_\eta(\rho)\kb{0}\right\|_1 \left\|(\1-\kb{0})\Phi^{\text{att}}_\eta(\rho)(\1-\kb{0})\right\|_1} \\&\le |\eta|\sqrt{\Tr\left(N\rho\right)}.
\end{align*}
The fourth term in \eqref{eq:AttTriangle}  admits the same bound by the same argument which leads us to
\begin{align*}
	\left\|\Phi^{\text{att}}_\eta(\rho) - P(\rho)\right\|_1  \le 2|\eta|\left(\Tr\left(N\rho\right) + \sqrt{\Tr\left(N\rho\right)}\right) \le 4|\eta|\, \Tr\left((N+\1)\rho\right).
\end{align*}
\end{proof}

For $t\ge 0$ we consider the parametrisation $\eta(t) = e^{-t}$ and define the \emph{attenuator semigroup} as $\left(\Phi^{\text{att}}_{\eta(t)}\right)_{t\ge 0}.$ In fact, using the established \eqref{eq:AttSemigroup} together with the fact that $\Phi^{\text{att}}_{\eta(0)} = \Phi^{\text{att}}_1=\id,$ we immediately see that this indeed defines an one-parameter semigroup. Furthermore, by Lemma~\ref{lem:AttenMixing} we see that it is strongly continuous and therefore has a generator $\cK_{\text{att}}$\footnote{It can be shown \cite[Lemma 13]{DePalmaGiovanneti_Attenuator_2016} that the generator of the attenuator semigroup can be explicitly written in terms of the canonoical creation and annihilation operators (see e.g.~\cite[Chapter 5.2]{BratteliRobinson_operator2_1997} or \cite[Chapter 8.3]{Teschl_mathematical_2014} for a definition and discussion) as $\cK_{\text{att}}(\rho) = 2a\rho a^* -N\rho - \rho N.$ Note that there a rescaled version of the attenuator semigroup is considered, namely $e^{t\cK_{\text{att}}/2}.$} with dense domain $\cD(\cK_{\text{att}})$ and can be written as 
\begin{align}
\label{eq:AttSemigroupDef}
\Phi^{\text{att}}_{\eta(t)} = e^{t\cK_{\text{att}}}.
\end{align} 

\medskip

\subsubsection{Mixing-, Zeno- and strong damping limits of the attenuator channel}
\smallskip

\noindent Note that from the semigroup property \eqref{eq:AttSemigroup} combined with strong continuity of the map $\eta\mapsto\Phi^{\text{att}}_\eta$ established in Lemma~\ref{lem:AttenMixing}, we see that for every $\eta\in\C$ with $|\eta|<1$ the corresponding attenuator channel is strongly power convergent. In fact we immediately get for every $x\in\cT(\cH)$ 
\begin{align}
\label{eq:AttStrongPowerConv1}
\lim_{n\to\infty} \left(\Phi^{\text{att}}_\eta\right)^n(x) =  \lim_{n\to\infty} \Phi^{\text{att}}_{\eta^n}(x) = \Phi^{\text{att}}_0(x) = P(x),
\end{align}
where we have again denoted the projection $P(x) = \Tr(x)\,\kb{0}.$

Similarly, we also get that the attenuator semigroup is strongly mixing as for every $x\in\cT(\cH)$ we have
\begin{align}
\label{eq:AttStrongMixing}
    \lim_{\gamma\to\infty} e^{\gamma\cK_{\text{att}}}(x) =  \lim_{\gamma\to\infty}\Phi^{\text{att}}_{e^{-\gamma}}(x) = P(x).
\end{align}

For $\rho\in\cT(\cH)_+$ which have a finite particle number, i.e.~$\Tr(N\rho)<\infty,$  \eqref{eq:AttMixingBound} in Lemma~\ref{lem:AttenMixing} gives even exponential speed of convergence for \eqref{eq:AttStrongPowerConv1} and \eqref{eq:AttStrongMixing}, i.e.
\begin{equation}
	\left\|\left(\Phi^{\text{att}}_\eta(\rho) \right)^n- P(\rho)\right\|_1 \le 4|\eta|^n\, \Tr\left((N+\1)\rho\right) \xrightarrow[n\to\infty]{}0.
\end{equation}
and 
\begin{equation}
\label{eq:AttStrongMixingBound}
	\left\|e^{\gamma\cK_{\text{att}}}(\rho) - P(\rho)\right\|_1 \le 4e^{-\gamma}\, \Tr\left((N+\1)\rho\right) \xrightarrow[\gamma\to\infty]{}0.
\end{equation}
Therefore, using the notation established in \eqref{eq:MpowerspeedZeno} and \eqref{eq:Kspeed} we see that we have the explicit bounds on the speed of convergence given as \begin{align}s_n(\rho)\le 4|\eta|^n\, \Tr\left((N+\1)\rho\right)\quad\quad\text{ and
}\quad\quad v_\gamma(\rho)\le 4e^{-\gamma}\, \Tr\left((N+\1)\rho\right).
\end{align}
We can now use these results together with Theorem~\ref{thm:StrongZenoGeneralQuant} and Theorem~\ref{thm:StrongDamping} to prove Zeno- and strong damping limits for the attenuator channel and semigroup. 
\begin{proposition}[Zeno- and strong damping limits for the attenuator]
\label{prop:AttZenoDamp}
    Let $\cL\in \cB(\cT(\cH))$ and  $x\in\cT(\cH)$. Then we have for all $\eta\in\C$ with $|\eta|<1$ and $t>0$ 
    \begin{align}
    \label{eq:AttStrongZenoLim}
        \lim_{n\to\infty} \left(\Phi^{\text{att}}_\eta e^{t\cL/n}\right)^n (x) = e^{tP\cL P}P(x) = e^{t\Tr(\cL(\kb{0}))}\, \Tr\left(x\right) \kb{0}
    \end{align}
    and
    \begin{align}
    \label{eq:AttStrongDampLim}
        \lim_{\gamma\to\infty} e^{t(\gamma \cK_{\text{att}}+\cL )}(x) = e^{tP\cL P}P(x) =  e^{t\Tr(\cL(\kb{0}))} \,\Tr\left(x\right) \kb{0},
    \end{align}
    where we denoted the projection $P x = \Tr(x) \kb{0}.$
    Moreover, we have the explicit bounds on the speed of convergence for all $n\ge 2$
    \begin{align}
    \label{eq:SpeedAttZeno}
      \nn&\left\| \left(\Phi^{\text{att}}_\eta e^{t\cL/n}\right)^n (x)  - e^{t\Tr(\cL(\kb{0})} \Tr\left(x\right) \kb{0}\right\|_1\\& \le \frac{C\log(n)}{n} \sum_{i=1}^4\left(\Tr\left((N+\1) x_i\right) + \Tr\left((N+\1) \cL(\kb{0})_i\right)\frac{\|x\|_1}{\|\cL\|}
\right),
    \end{align}
    and for all $\gamma\ge 2$
    \begin{align}
    \label{eq:SpeedAttDamping}
      \nn&\left\| e^{t(\gamma \cK_{\text{att}}+\cL )}(x)  - e^{t\Tr(\cL(\kb{0})} \Tr\left(x\right) \kb{0}\right\|_1\\& \le \frac{C\log(\gamma)}{\gamma} \sum_{i=1}^4\left(\Tr\left((N+\1) x_i\right) + \Tr\left((N+\1) \cL(\kb{0})_i\right)\frac{\|x\|_1}{\|\cL\|}
\right),
    \end{align}
    for all $\gamma\ge e$ and some $C\ge 1$ independent of $\gamma$ and $n$ respectively and 
    where $x=x_1-x_2 + i(x_3-x_4)$ and $\cL(\kb{0})= \cL(\kb{0})_1-\cL(\kb{0})_2 +i\left(\cL(\kb{0})_3-\cL(\kb{0})_4\right)$ denote any decomposition into a sum of four positive semi-definite operators.\footnote{For all $x\in\cB(\cH)$ such decompositions always exists, e.g.~using \emph{real and imaganiry parts} defined by $\Re(x) \coloneqq (x+x^*)/2$ and $\Im(x) \coloneqq (x-x^*)/2$ and then their positive and negative parts respectively, i.e.~$x_1 = [\Re(x)]_+,$ $x_2 = [\Re(x)]_-,$ $x_3 = [\Im(x)]_+$ and $x_4 = [\Im(x)]_-.$}
\end{proposition}
\begin{remark}
\label{rem:AttZeno}
Note that if $\left(e^{t(\gamma\cK_{\text{att}}+\cL)}\right)_{t\ge 0}$ is a semigroup of quantum channels and hence hermiticity- and trace preserving, we see that $\cL$ must be hermiticity preserving as well and every operator in its image must be traceless. Therefore, in this case and for a state $\rho$ with finite particle number, i.e.~$\Tr(N\rho)<\infty,$ the bounds \eqref{eq:SpeedAttZeno} and \eqref{eq:SpeedAttDamping} simplify to
\begin{align*}
\left\| \left(\Phi^{\text{att}}_\eta e^{t\cL/n}\right)^n(\rho) - \kb{0}\right\|_1\le \frac{C\log(n)}{n} \left(\Tr\left(N +\1)\rho\right) + \frac{\Tr\left((N+\1) ([\cL(\kb{0})]_+ +[\cL(\kb{0})]_-)\right)}{\|\cL\|}\right)
\end{align*}
and 
\begin{align*}
\left\| e^{t(\gamma\cK_{\text{att}}+\cL)}(\rho) - \kb{0}\right\|_1 \le \frac{C\log(\gamma)}{\gamma} \left(\Tr\left(N +\1)\rho\right) + \frac{\Tr\left((N+\1) ([\cL(\kb{0})]_+ +[\cL(\kb{0})]_-)\right)}{\|\cL\|}\right).
\end{align*}
Therefore, under the assumption that $\cL$ does not generate infinite number of particles from the \emph{vacuum state} $\kb{0},$ with which we mean precisely $\Tr\left(N [\cL(\kb{0})]_+ \right),\Tr\left(N[\cL(\kb{0})]_-\right)<\infty$, we see that the attenuator converges in the Zeno- and strong damping regime to the vacuum with speed of convergence being $\mathcal{O}(\log(n)/n)$ and $\mathcal{O}(\log(\gamma)/\gamma)$ respectively.
\end{remark}
\begin{proof}[Proof of Proposition~\ref{prop:AttZenoDamp}]
The strong Zeno- and damping limits, \eqref{eq:AttStrongZenoLim} and \eqref{eq:AttStrongDampLim}, directly follow by using \eqref{eq:AttStrongPowerConv1} combined with Theorem~\ref{thm:StrongZenoGeneralQuant} or \eqref{eq:AttStrongMixing} combined with Theorem~\ref{thm:StrongDamping} respectively.

We now proceed to prove the bound on the speed of convergence \eqref{eq:SpeedAttZeno} and \eqref{eq:SpeedAttDamping}. We will focus on \eqref{eq:SpeedAttDamping} as \eqref{eq:SpeedAttZeno} follows using the respective statement of Theorem~\ref{thm:StrongZenoGeneralQuant}. Using the notation of \eqref{eq:StrongDampSpeed}, we see by \eqref{eq:AttStrongMixingBound} together with the decomposition of $x\in\cT(\cH)$ into a sum of four positive semi-definite operators, $x= x_1-x_2 +i\left(x_3-x_4\right)$, and the triangle inequality that
\begin{align*}
    v_\gamma(x) = \sup_{\gamma'\ge\gamma}\left\|\left(e^{\gamma \cK_{\text{att}}} - P\right)x\right\|_1 \ \le 4e^{-\gamma}\sum_{i=1}^4\Tr\left((N+\1)x_i\right)
\end{align*}
 Using this together with \eqref{eq:StrongDampSpeed} in Theorem~\ref{thm:StrongDamping} gives for the constant sequence $N_l = N_1$ for all $l\in\N$ and $N_1\in\N$ to be determined later and some $C\ge 1$
\begin{align*}
&\left\|e^{t(\gamma\cK_{\text{att}} + \cL)}x -  e^{t\Tr(\cL(\kb{0}))}\Tr\left(x\right) \kb{0}\right\|_1 \le C \sup_{l\in\N}\left(\|t\cL\|^{-l+1}v_{N_l}\left((t\cL P)^{l-1}x\right) +\frac{N_l}{\floor{\gamma}}\|x\|_1\right)\\&\le 4C \left(e^{-N_1}\sum_{i=1}^4\left(\,\Tr\left((N+\1) x_i\right) + \Tr\left((N+\1) \cL(\kb{0})_i\right)\frac{\|x\|_1}{\|\cL\|}
\right)+ \frac{N_1\|x\|_1}{\floor{\gamma}}\right), 
\end{align*}
where for the second inequality we have used that $\left(\cL P\right)^{l-1}x=x$ for $l=1$ and  
\begin{align*}
     \left(\cL P\right)^{l-1}x = 
      \cL(\kb{0})\,\Tr\left(\cL(\kb{0})\right)^{l-2} \Tr(x),
\end{align*}
for $l\ge 2$ together with the fact that $|\Tr\left(\cL(\kb{0})\right)|\le \|\cL\|.$
For $\gamma\ge 2$ we now choose $N_1 = \floor{\log(\gamma)}$ gives
\begin{align*}
&\left\|e^{t(\gamma\cK_{\text{att}} + \cL)}x -  e^{t\Tr(\cL(\kb{0}))}\Tr\left(x\right) \kb{0}\right\|_1 \\&\le 4C \left(\frac{e}{\gamma}\sum_{i=1}^4\left(\,\Tr\left((N+\1) x_i\right) + \Tr\left((N+\1) \cL(\kb{0})_i\right)\frac{\|x\|_1}{\|\cL\|}
\right)+ \frac{2\log(\gamma)\|x\|_1}{\floor{\gamma}}\right).
\end{align*}
Using $\|x\|_1 \le \sum_{i=1}^4\Tr(x_i)$ and $\frac{1}{\floor{\gamma}} \le \frac{2}{\gamma}$ for all $\gamma\ge 2$, this gives the desired bound \eqref{eq:SpeedAttDamping} after renaming the constant $C.$
\end{proof}

\section{Proof of Theorem~\ref{thm:StrongBinoFormula}}
\label{sec:ProofOfBino}
We now proceed to give the proof of Theorem~\ref{thm:StrongBinoFormula} showing the asymptotic behaviour of the generalised binomial product
\begin{align}
\label{eq:GenProdProofSec}
    \left(M_r +\frac{\cL_{n,r}}{n}\right)^n
\end{align}
defined in \eqref{eq:GenOpProd} as $n\to\infty$. For that we establish a quantitative version of the perturbation series technique developed in \cite[Theorem 2]{BeckerDattaSalz_Zeno_2021}, enabling us to provide explicit bounds on the speed of convergence of the generalised binomial product. 

\begin{figure}[t]
	 	\centering
	 		\includegraphics[width=0.55\linewidth]{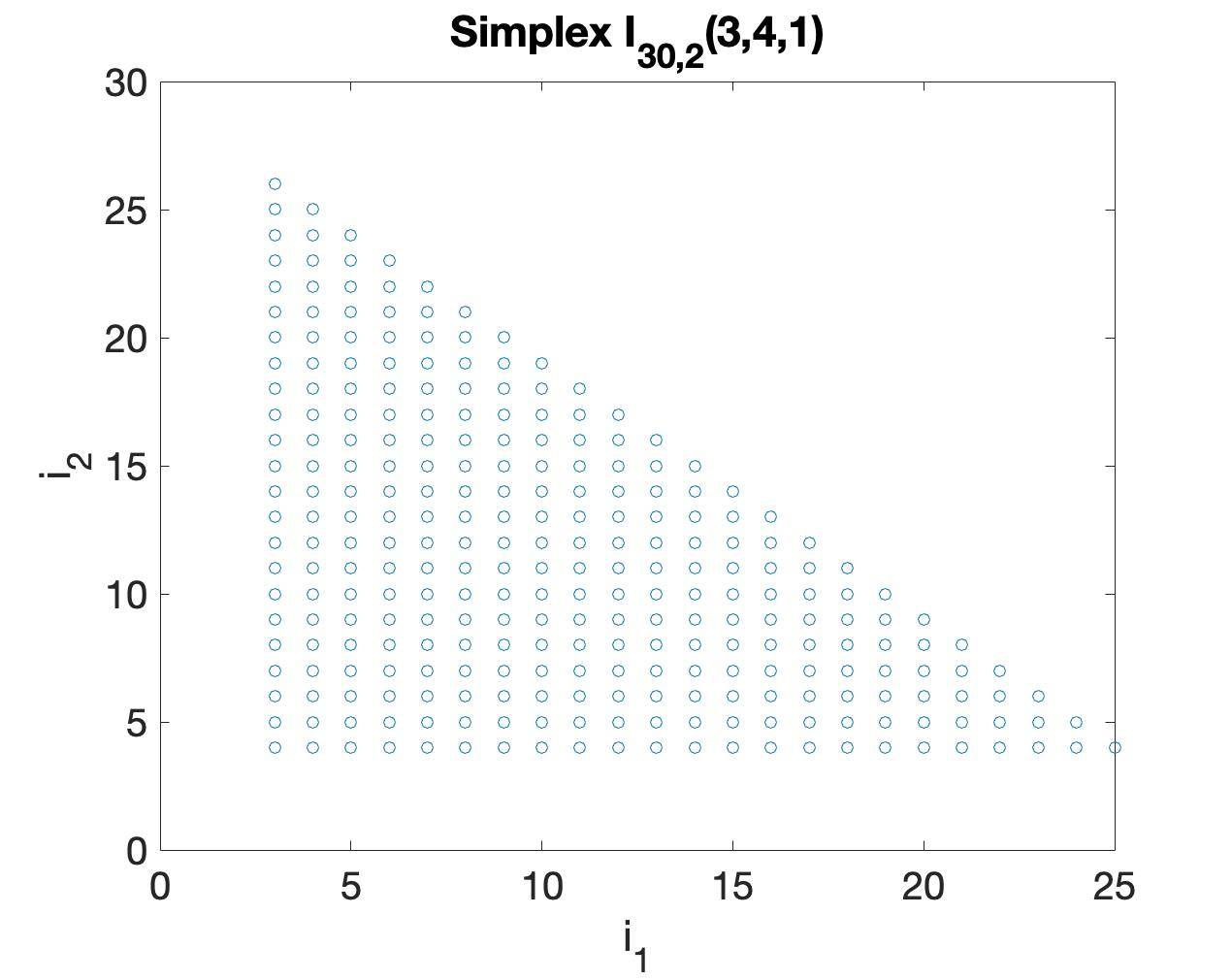}
	 	\caption{Sets $I_{n,k}(\mathbf N)$, defined in \eqref{eq:Ink} take the form of discrete simplexes.}
	 	\label{fig:simplex}
	 \end{figure}

We define for all $n,k \in \mathbb N$ the discrete simplex
$$\Delta_{\text{disc}}^k(n) \coloneqq \left\{(i_1,\dots,i_{k})\in\mathbb{N}_0^{k}~\Big|\,\sum_{l = 1}^{k} i_l\le n -k\right\}.$$
 Moreover, we consider the following subsets of the discrete simplex for all $\mathbf N:=(N_1,\cdots,N_{k+1})\in\N^{k+1}$ as
\begin{equation*}
    \begin{split}
    \label{eq:Ink}
I_{n,k}(\mathbf N) := \left\{i\in\N_0^k\,\Big|\,i_l\ge N_l\,\,\forall l\in[k], \,\sum_{l=1}^k i_l \le n- k-N_{k+1}\right\}.
    \end{split}
\end{equation*}
See Figure~\ref{fig:simplex} for an illustration.
Note, that by definition we have $I_{n,k}(\mathbf{N})\subseteq I_{n,k}(0,\cdots,0)=\Delta_{\text{disc}}^k(n).$
For $i=(i_1,\dots,i_k)\in\Delta^k_{\text{disc}}(n)$ we use the notation $i_{k+1} = n - k-\sum_{l=1}^k i_l$. Moreover, for operators $A_1,\cdots,A_k\in\cB(X)$ we denote the non-commutative product by
\begin{equation}
\Pi_{l=k}^1 A_l=A_kA_{k-1}\cdots A_1.
\end{equation}
For the proof of Theorem~\ref{thm:StrongBinoFormula}, we split the generalised binomial product into a sum consisting of terms corresponding to different powers of $1/n$ of the form 
\begin{align}
\label{eq:ExpandOpis}
    \frac{1}{n^k}\sum_{i \in \Delta_{\text{disc}}^k(n)} \Big(\Pi_{m=k+1}^2M_r^{i_m}\cL_{n,r}\Big)M_r^{i_1}.
\end{align}
The main technical tool is then to show strong convergence of these operators in Lemma~\ref{lem:ZenoWorkhorse}. For that we first analyse the asymptotic behaviour of the cardinalities of the discrete simplices $\Delta_{\text{disc}}^k(n)$ and $I_{n,k}(\mathbf{N})$
as $n \to \infty$ in Lemma~\ref{lem:cardinality} below. To also be able to provide the bound on the convergence speed \eqref{eq:SpeedBino}, it is essential to have good quantitative control on these asymptotics.

The idea to prove convergence of the operators \eqref{eq:ExpandOpis} can then be sketched as follows (see proof of Lemma~\ref{lem:ZenoWorkhorse} for details): Pick $\mathbf{N} =(N_1,\cdots,N_{k+1})$ with all components being large, e.g.~we can for now think of a $n$ dependent $\mathbf{N}(n)$ with all components going to infinity as $n\to\infy.$ Then for $(i_1(n),\cdots,i_k(n))\in I_{n,k}(\mathbf{N}(n))$ we see by the assumptions of Theorem~\ref{thm:StrongBinoFormula} for $x\in X$
\begin{align*}
  \lim_{n\to\infty} \Big(\Pi_{m=k+1}^2M_r^{i_m(n)}\cL_{n,r}\Big)M_r^{i_1(n)} x = \left(P\cL_r P\right)^kx.
\end{align*} 
We then use that the remainder of summands in \eqref{eq:ExpandOpis} with $i\in \Delta_{\text{disc}}^k(n)\setminus I_{n,k}(\mathbf{N}(n))$ are negligible by \eqref{eq:SimplexBound2} in Lemmma~\ref{lem:cardinality} (provided the components of $\mathbf{N}(n)$ do not grow too drasticially as $n\to\infty$). Combining this with \eqref{eq:CardLimitDelta} in Lemma~\ref{lem:cardinality} gives
\begin{align}
\label{eq:ExpandOpisLimi}
    \lim_{n\to\infty}\frac{1}{n^k}\sum_{i \in \Delta_{\text{disc}}^k(n)} \Big(\Pi_{m=k+1}^2M_r^{i_m}\cL_{n,r}\Big)M_r^{i_1}x  = \frac{\left(P\cL_r P\right)^k}{k!}x.
\end{align}
The last step for proving Theorem~\ref{thm:StrongBinoFormula} is then to sum the operators \eqref{eq:ExpandOpis} over $k\in[n]$ and recover the exponential series as $n\to\infty$ from \eqref{eq:ExpandOpisLimi}. 

We now give all the technical details for the  Lemmas~\ref{lem:cardinality} and~\ref{lem:ZenoWorkhorse} as well as the proof of Theorem~\ref{thm:StrongBinoFormula}. 

\begin{lemma}
\label{lem:cardinality}
	For all $k\in\N$ we have 
 \begin{align}
    \label{eq:CardLimitDelta}
    \lim_{n\to\infty}\frac{|\Delta^k_{\text{disc}}(n)|}{n^k} = \frac{1}{k!},
 \end{align}
 with bound on the speed of convergence 
\begin{align}
\left|\frac{|\Delta^k_{\text{disc}}(n)|}{n^k} - \frac{1}{k!}\right| \le \frac{2^k}{(k-1)!\,n},
\end{align}
for all $n\ge k.$
Moreover, for all $\mathbf N:=(N_1,\cdots,N_{k+1})\in\N^{k+1}$ and $n\ge k$ we have
\begin{align}
\label{eq:SimplexBound2}
        \left|\frac{|\Delta^k_{\text{disc}}(n)| - |I_{n,k}(\mathbf N)|}{n^k}\right| \le \frac{1}{(k-1)!\, n}\sum_{l=1}^{k+1}N_l.
\end{align}

\end{lemma}
\begin{proof}
	To simplify notation we introduce $m=n-k.$ First note that
	\begin{align*}
	\frac{|\Delta^k_{\text{disc}}(n)|}{n^k} = \frac{1}{n^k}\sum_{(i_1,\dots,i_k)\in\Delta_{\text{disc}}^k(n)}1 =\frac{1}{n^k} \sum_{i_1=0}^{m}\sum_{i_2=0}^{m-i_1}\cdots\sum_{i_{k-1}=0}^{m-\sum_{l=1}^{k-2}i_l}\sum_{i_k=0}^{m-\sum_{l=1}^{k-1}i_l}1.
	\end{align*}
	If we denote by $$\Delta^k = \left\{(t_1,\dots,t_{k})\in\mathbb{R}^{k}~\Big|~\sum_{i = 1}^{k} t_i \le 1 \text{ and } t_i \ge 0 \text{ for all } i\right\}$$ the $k$-simplex, then we obtain as a limiting expression, for the limit $n\to \infty$, the volume of the $k$-simplex
\begin{align*}
\lim_{n\to\infty}\frac{|\Delta^k_{\text{disc}}(n)|}{n^k} &= \int_0^1\int^{1-t_1}_0\int_0^{1-t_1-t_2}\cdots\int_0^{1-\sum_{l=1}^{k-1} t_{l}} 1\,d t_k d t_{k-1}\cdots d t_1 \\ &=\int_{\Delta^k} 1 \ dt =\frac{1}{k!}\,. 
\end{align*} 
In the following we prove a bound on the speed of convergence for the above limit. Denoting the $k$-dimensional box with side length $1/n$ and base point $(i_1,\cdots,i_k)\in\N^k_0$ by
\begin{align*}
B(i_1,\dots,i_k) = \left[\frac{i_1}{n},\frac{i_1+1}{n}\right[\times\cdots\times\left[\frac{i_k}{n},\frac{i_k+1}{n}\right[
\end{align*}
with volume 
\begin{align*}
   \text{vol}\big(B(i_1,\dots,i_k)\big) = \int_{B(i_1,\dots,i_k)} 1\ dt = \frac{1}{n^k}.
\end{align*}
From the definition of $\Delta^k$ it is easy to see that we have the inclusions
\begin{align}
\label{eq:SimplexInclusion}
 \left(1-\frac{k}{n}\right)\Delta^k\subseteq \bigcup_{(i_1,\cdots,i_k)\in\Delta^{k}_{\text{disc}}(n)}B(i_1,\dots,i_k) \subseteq\Delta^k.
\end{align}
Therefore, using that for $(i_1,\cdots,i_k)\neq (i'_1,\cdots,i'_k)$ the sets $B(i_1,\dots,i_k)$ and $B(i'_1,\dots,i'_k)$ are disjoint, we see
\begin{align}
\label{eq:SimplUniUpperBound} \nn\frac{|\Delta^k_{\text{disc}}(n)|}{n^k} &= \sum_{(i_1,\dots,i_k)\in\Delta_{\text{disc}}(n)}\text{vol}\big(B(i_1,\dots,i_k)\big) \\&= \text{vol}\left(\bigcup_{(i_1,\dots,i_k)\in\Delta_{\text{disc}}(n)}B(i_1,\dots,i_k)\right)\le \text{vol}(\Delta^k) = \frac{1}{k!}
\end{align}
and by the same argument
\begin{align*}
\frac{|\Delta^k_{\text{disc}}(n)|}{n^k}\ge\text{vol}\left(\left(1-\frac{k}{n}\right)\Delta^k\right).
\end{align*}
Hence, this gives
\begin{align*}
\label{eq:SomeCombiFombi}
\nn\left|\frac{|\Delta^k_{\text{disc}}(n)|}{n^k} -  \text{vol}(\Delta^k) \right| &\le  \text{vol}(\Delta^k)-\text{vol}\left(\left(1-\frac{k}{n}\right)\Delta^k\right) =  \left(1-\left(1-\frac{k}{n}\right)^k\right)\text{vol}(\Delta^k)\\&\le \frac{2^kk}{n}\text{vol}(\Delta^k) =\frac{2^k}{(k-1)!\,n},
\end{align*}
where for the second inequality we explicitly used $n\ge k.$

We now proceed to prove the bound \eqref{eq:SimplexBound2}. Without loss of generality assume $\sum_{l=1}^{k+1}N_l \le n-k$ as otherwise $I_{n,k}(\mathbf N)=\emptyset$ and hence the bound follows from the established \eqref{eq:SimplUniUpperBound}.
Denoting again $m=n-k$, we see
\begin{align*}
 &|\Delta^k_{\text{disc}}(n)| - |I_{n,k}(\mathbf N)| = \sum_{i_1=0}^{m}\sum_{i_2=0}^{m-i_1}\cdots\sum_{i_{k-1}=0}^{m-\sum_{l=1}^{k-2}i_l}\sum_{i_k=0}^{m-\sum_{l=1}^{k-1}i_l}1 -\sum_{i_1=N_1}^{m}\sum_{i_2=N_2}^{m-i_1}\cdots\sum_{i_{k-1}=N_{k-1}}^{m-\sum_{l=1}^{k-2}i_l}\sum_{i_k=N_k}^{m-N_{k+1}-\sum_{l=1}^{k-1}i_l}1 \\&\quad =\sum_{i_1=0}^{m}\sum_{i_2=0}^{m-i_1}\cdots\sum_{i_{k-1}=0}^{m-\sum_{l=1}^{k-2}i_l}\left(N_k+N_{k+1}\right)\\&\quad\quad\ +\left(\sum_{i_1=0}^{m}\sum_{i_2=0}^{m-i_1}\cdots\sum_{i_{k-1}=0}^{m-\sum_{l=1}^{k-2}i_l} -\sum_{i_1=N_1}^{m}\sum_{i_2=N_2}^{m-i_1}\cdots\sum_{i_{k-1}=N_{k-1}}^{m-\sum_{l=1}^{k-2}i_l}\right)\sum_{i_k=N_k}^{m-N_{k+1}-\sum_{l=1}^{k-1}i_l}1\\&\quad \le |\Delta^{k-1}_{\text{disc}}(n)|\sum_{l=1}^{k+1} N_l,
\end{align*}
where for the last step, we have used for the first term before the inequality that
\begin{align*}
\sum_{i_1=0}^{m}\sum_{i_2=0}^{m-i_1}\cdots\sum_{i_{k-1}=0}^{m-\sum_{l=1}^{k-2}i_l}\left(N_k+N_{k+1}\right) = |\Delta^{k-1}_{\text{disc}}(n)|\left(N_k+N_{k+1}\right) 
\end{align*}
and for the second term we iteratively used
\begin{align*}
&\left(\sum_{i_1=0}^{m}\sum_{i_2=0}^{m-i_1}\cdots\sum_{i_{k-1}=0}^{m-\sum_{l=1}^{k-2}i_l} -\sum_{i_1=N_1}^{m}\sum_{i_2=N_2}^{m-i_1}\cdots\sum_{i_{k-1}=N_{k-1}}^{m-\sum_{l=1}^{k-2}i_l}\right)\sum_{i_k=N_k}^{m-N_{k+1}-\sum_{l=1}^{k-1}i_l}1 \\&\quad\le \sum_{i_1=0}^{m}\sum_{i_2=0}^{m-i_1}\cdots\sum_{i_{k-2}=0}^{m-\sum_{l=1}^{k-3}i_l}\sum_{i_k=0}^{m-\sum_{l=1}^{k-2}i_l} N_{k-1} \\&\quad\quad\ + \left(\sum_{i_1=0}^{m}\sum_{i_2=0}^{m-i_1}\cdots\sum_{i_{k-2}=0}^{m-\sum_{l=1}^{k-3}i_l} -\sum_{i_1=N_1}^{m}\sum_{i_2=N_2}^{m-i_1}\cdots\sum_{i_{k-2}=N_{k-2}}^{m-\sum_{l=1}^{k-3}i_l}\right)\sum_{i_{k-1}=N_{k-1}}^{m-\sum_{l=1}^{k-2}i_l}\sum_{i_k=N_k}^{m-N_{k+1}-\sum_{l=1}^{k-1}i_l}1 \\&\le\cdots\le |\Delta^{k-1}_{\text{disc}}(n)|\sum_{l=1}^{k-1}N_l.
\end{align*}
Hence, using that by the same argument as for \eqref{eq:SimplUniUpperBound} we have 
\begin{align*}
    \frac{|\Delta^{k-1}_{\text{disc}}(n)|}{n^{k-1}} \le \text{vol}\left(\Delta^{k-1}\right) =  \frac{1}{(k-1)!},
\end{align*}
where again we have used $n\ge k,$ we conclude
\begin{align*}
    \frac{|\Delta^k_{\text{disc}}(n)| - |I_{n,k}(\mathbf N)|}{n^k} \le \frac{1}{(k-1)!\, n}\sum_{l=1}^{k+1}N_l.
\end{align*}
\end{proof}  
We are now ready to analyse the asymptotic behaviour of the operators \eqref{eq:ExpandOpis}.
\begin{lemma}
	\label{lem:ZenoWorkhorse} 
	Let $R$ be a compact interval and $(\cL_{n,r})_{n\in\N}\subseteq\cB(X)$ be operators which, uniformly in $r\in R$, converge to some operator $\cL_r$. Furthermore, assume that $r\mapsto \cL_r$ is strongly continuous.\\
Let $R'\subseteq R$ and $\left(M_r\right)_{r\in R'}\subseteq \cB(X)$ be contractions such that for all $x\in X$ 
 \begin{equation}
     \label{eq:MrstrongpowerconvergenctThm}
     \lim_{n\to\infty} M^n_rx = Px,
 \end{equation}
uniformly in $r\in R'$ and for some $P\in\cB(X).$   
Then for all $k\in\N$ 
\begin{equation}
\label{eq:limitpimit}
\lim_{n\to\infty}\frac{1}{n^k}\sum_{i\in \Delta_{\text{disc}}^k(n)} \Big(\Pi_{m=k+1}^2M_r^{i_m}\cL_{n,r}\Big)M_r^{i_1}x=\frac{(P\cL_r P)^k}{k!} x 
\end{equation}
    uniformly in $r\in R'.$ More precisely,  using the notation established in \eqref{eq:LConv} and \eqref{eq:Mpowerspeed}, we have for all $n\ge k$  and $\mathbf{N}=(N_1,\cdots,N_{k+1})\in\N^{k+1}$ the bound
    \begin{equation}
    \begin{split}
		\label{eq:limitpimitBino}
		&\left\|\left(\frac{1}{n^k}\sum_{i \in \Delta_{\text{disc}}^k(n)} \Big(\Pi_{m=k+1}^2M_r^{i_m}\cL_{n,r}\Big)M_r^{i_1}  -\frac{(P\cL_r P)^k}{k!} \right)x\right\| \le\frac{kC^k}{(k-1)!}\Big(\,s'_n\|x\| + s^{\max}_n\left(x,\cL_r,\mathbf{N}\right)\Big),
  \end{split}
	\end{equation} 
 for some $C\ge1$ independent of $n$ and $k,$ 
 where we denoted \begin{equation}
s^{\max}_n(x,\cL_r,\mathbf{N})=\max_{l\in[k+1]}\left(\left(\sup_{\tilde r\in R}\|\cL_{\tilde r}\|\right)^{-l+1}s_{N_l}\left((\cL_rP)^{l-1}x\right) +\frac{N_l}{n}\|x\|\right).
 \end{equation}
\end{lemma}
\begin{proof}
Since $r\mapsto \cL_r$ is strongly continuous on $R$, which is a compact set, we have for all $x\in X$ that $\sup_{r\in R}\|\cL_rx\|<\infty$ and hence by the uniform boundedness principle \cite[Theorem III.9]{ReedSimon_FunctionalAnalysis_1976} also $\sup_{r\in R}\|\cL_r\|<\infty.$ Therefore, since $\lim_{n\to\infty}\cL_{n,r} = \cL_r$ uniformly in $r\in R$, we also have $C=\sup_{n\in\N,r\in R}\left\|\cL_{n,r}\right\|<\infty.$ Let $r\in R'\subseteq R$ and using that $M_r$ is a contraction, we have for all $l\in[k]$ and $i=(i_1,\cdots,i_k)\in\Delta_{\text{disc}}^k(n)$ 
\begin{align*}
		 \left\|\Pi_{m=k+1}^{l+1}(M_r^{i_m}\cL_{n,r})\right\| \le C\left\|\Pi_{m=k}^{l+1}(M_r^{i_m}\cL_{n,r})\right\| \le \cdots \le C^{k-l+1}. 
	\end{align*} 
Hence, for all $y\in X$ using the notion of  \eqref{eq:Mpowerspeed}, we have
	\begin{equation}
 \begin{split}
	    \label{eq:MProdBound}
		\left\| \Pi_{m=k+1}^{l+1}(M_r^{i_m}\cL_{n,r})(M_r^{i_{l}} -P)(y)\right\|  & \le\left\|\Pi_{m=k+1}^{l+1}(M_r^{i_m}\cL_{n,r})\right\| \left\|(M_r^{i_l} -P)(y)\right\|\\& \le C^{k-l+1} s_{i_l}(y).
  \end{split}
	\end{equation}
	Moreover, using the fact that $P$ as the strong limit of the contractions $(M^n_r)_{n\in\N}$ is a contraction itself and the notion of \eqref{eq:LConv}, we see by similar argument 
	\begin{equation}
 \begin{split}
 \label{eq:LProdBound}
		&\left\| \Pi_{m=k+1}^{l+1}(M^{i_m}_r\cL_{n,r})M^{i_l}( \cL_{n,r}-\cL_r)(P\cL_rP)^{l-1}  \right\|  \\& \le 	\left\| \Pi_{m=k+1}^{l+1}(M^{i_m}_r\cL_{n,r})M^{i_l}\right\|\left\|( \cL_{n,r}-\cL_r)\right\| \left\|(P\cL_rP)^{l-1}  \right\| \le C^{k} s'_{n}. 
  \end{split}
	\end{equation} 	
	 We combine above inequalities using the triangle inequality to bound the quantity of interest. For that, first note that by using \eqref{eq:MProdBound} we have
	\begin{align*}
		&\left\|\Big(\Pi_{m=k+1}^{2}(M_r^{i_m}\cL_{n,r})  M_r^{i_1} -\big(P\cL_r P\big)^{k}\Big)x\right\| \\&\le  \left\|\Pi_{m=k+1}^{2}(M_r^{i_m}\cL_{n,r})  (M_r^{i_1} - P)x\right\| \\&\quad + \left\|\Big(\Pi^2_{m=k+1}(M_r^{i_m}\cL_{n,r}) P -\big(P\cL_r P\big)^{k}\Big)x\right\| \\&\le C^k s_{i_1}(x)+  \left\|\Big(\Pi^2_{m=k+1}(M_r^{i_m}\cL_{n,r} ) P -\big(P\cL_r P\big)^{k}\Big)x\right\|.
	\end{align*}
	Estimating the second term in the last line by using \eqref{eq:LProdBound} gives
	\begin{equation*}
		\begin{split}
			&\left\|\Big( \Pi^2_{m=k+1}(M_r^{i_m}\cL_{n,r} ) P -\big(P\cL_r P\big)^{k}\Big)x\right\|\\
			&\le \left\|\Big(\Pi^3_{m=k+1}(M_r^{i_m}\cL_{n,r} )M_r^{i_2}(\cL_{n,r} - \cL_r )\Big)\right\|\|x\| \\
			&\quad + \left\|\Big( \Pi^3_{m=k+1}(M_r^{i_m}\cL_{n,r}) M_r^{i_2}\cL_r P -\big(P\cL_r P\big)^{k}\Big)x\right\| \\&\le C^k s'_n\|x\| + \left\|\Big( \Pi^3_{m=k+1}(M_r^{i_m}\cL_{n,r}) M_r^{i_2}\cL_r P -\big(P\cL_r P\big)^{k}\Big)x\right\| .
		\end{split}
	\end{equation*}
	By iterating these estimates, we obtain
	\begin{equation}
		\label{eq:Estimates}
		\begin{split}
			\left\|\Big( \Pi_{m=k+1}^2(M_r^{i_m}\cL_{n,r} ) -\big(P\cL_r P\big)^{k}\Big)x\right\|&\le C^kk s'_n\|x\| +  \sum_{l=1}^{k+1}C^{k-l+1}s_{i_l}\left((\cL_rP)^{l-1}x\right)\\&\le C^k \left(k s'_n\|x\| +  \sum_{l=1}^{k+1}\left(\sup_{\tilde r\in R}\|\cL_{\tilde r}\|\right)^{-l+1}s_{i_l}\left((\cL_rP)^{l-1}x\right)\right).
		\end{split}
	\end{equation}
	 Moreover, clearly we also have the bound
 \begin{equation*}
		\begin{split}
		\left\|\Big( \Pi_{m=k+1}^2(M_r^{i_m}\cL_{n,r} ) -\big(P\cL_r P\big)^{k}\Big)x\right\|
			\le 2C^k\|x\|.
		\end{split}
	\end{equation*}
	Therefore, for every $0\neq\mathbf{N}=(N_1,\dots,N_{k+1})\in\N_0^k$ we see 
 \begin{align*}
		&\left\|\left(\,\frac{1}{n^k}\sum_{i \in \Delta_{\text{disc}}^k(n)} \Pi_{m=k+1}^{2}(M_r^{i_m}\cL_{n,r})  M_r^{i_1} - \frac{(P\cL_r P)^k}{k!}\right) x\right\| \\
		&\le\frac{1}{n^k}\sum_{i \in \Delta_{\text{disc}}^k(n)}\left\| \left(\Pi_{m=k+1}^{2}(M_r^{i_m}\cL_{n,r})  M_r^{i_1} - (P\cL_r P)^k\right) x\right\|\\&\quad +\left|\frac{|\Delta^k_{\text{disc}}(n)|}{n^k}-\frac{1}{k!}\right|\|(P\cL_r P)^kx\| \\&\le \frac{1}{n^k}\sum_{i \in I_{n,k}(\mathbf{N})}\left\| \left(\Pi_{m=k+1}^{2}(M_r^{i_m}\cL_{n,r})  M_r^{i_1} - (P\cL_r P)^k\right) x\right\|\\&\quad +2C^k\|x\|\left(\,\left|\frac{|\Delta^k_{\text{disc}}(n)|- |I_{n,k}(\mathbf{N})|}{n^k}\right|+\left|\frac{|\Delta^k_{\text{disc}}(n)|}{n^k}-\frac{1}{k!}\right|\right)\\&\le \frac{C^k}{k!}\left(k\,s'_n\|x\| + \sum_{l=1}^{k+1} \left(\sup_{\tilde r\in R}\|\cL_{\tilde r}\|\right)^{-l+1}s_{N_l}\left((\cL_rP)^{l-1}x\right) \right) + \frac{2(2C)^k}{(k-1)!\, n}\sum_{l=1}^{k+1}N_l.
 	\end{align*}
    where in the last line we have used \eqref{eq:Estimates} together with the fact that $n\mapsto s_n(y)$ is monotonically decreasing for every $y\in X$, the bounds presented in Lemma~\ref{lem:cardinality} and the fact that
    \begin{align*}
        \frac{|I_{n,k}(\mathbf{N})|}{n^k} \le \frac{|\Delta^k_{\text{disc}}(n)|}{n^k} \le\frac{1}{k!},
    \end{align*}
    which follows from \eqref{eq:SimplUniUpperBound} and the fact that $I_{n,k}(\mathbf{N})\subseteq\Delta^k_{\text{disc}}(n).$
    Using now \begin{align}
    \label{eq:FinalIneLemma}
\nn &\frac{C^k}{k!}\left(k\,s'_n\|x\| + \sum_{l=1}^{k+1} \left(\sup_{\tilde r\in R}\|\cL_{\tilde r}\|\right)^{-l+1}s_{N_l}\left((\cL_rP)^{l-1}x\right) \right) + \frac{2(2C)^k}{(k-1)!\, n}\sum_{l=1}^{k+1}N_l\\& \le \frac{8k(2C)^k}{(k-1)!}\left(\,s'_n\|x\| + \max_{l\in[k+1]}\left(\left(\sup_{\tilde r\in R}\|\cL_{\tilde r}\|\right)^{-l+1}s_{N_l}\left((\cL_rP)^{l-1}x\right) +\frac{N_l}{n}\|x\|\right)\right)
    \end{align}
    and renaming the constant $C$ gives the bound in \eqref{eq:limitpimitBino}.
 	Finally, we convince ourselves that from \eqref{eq:limitpimitBino} we can conclude \eqref{eq:limitpimit}. Let $\eps>0$. Since for all $l\in[k+1]$ we have $\lim_{n\to\infty}s_n((\cL_r P)^{l-1}x)=0$ and furthermore $n\mapsto s_n((\cL_r P)^{l-1}x)$ being monotonically decreasing and $r\mapsto s_n((\cL_r P)^{l-1}x)$ being continuous on the compact set $R$, we can pick by Dini's Theorem \cite[Theorem 7.13]{Rudin_Analysis_1976} for all $l\in[k+1]$ a $N_l\in\N$ big enough such that $\sup_{r\in R}s_{N_l}\left((\cL_rP)^{l-1}x\right)<\eps$  and then $N\in\N$ big enough such that $s'_n<\eps$ and $N_l/n<\eps$ for all $n\ge N.$ Hence, all terms in \eqref{eq:FinalIneLemma} are, uniformly in $r\in R$, upper bounded by some $k$ dependent constant times $\eps,$ 
    which concludes the proof.
 \end{proof}
 
 We can now give the proof of Theorem~\ref{thm:StrongBinoFormula}:

 \begin{proof}[Proof of Theorem~\ref{thm:StrongBinoFormula}]
	As above we use the convention $i_{k+1} = n -k-\sum_{l=1}^k i_l$ and write $\Pi_{m={k+1}}^{2}(M_r^{i_m}\cL_{n,r}) = M_r^{i_{k+1}}\cL_{n,r}M_r^{i_{k}}\cL_{n,r}\cdots M_r^{i_{2}}\cL_{n,r}.$
	We expand for all $x\in X$ and $r\in R'\subseteq R$
	\begin{align*}
	\left(M + \frac{\cL_{n,r}}{n}\right)^n x & = M^nx + \sum_{k=1}^n\frac{1}{n^k}\sum_{i \in \Delta_{\text{disc}}^k(n)} \Big(\Pi_{m=k+1}^2M^{i_m}\cL_{n,r}\Big)M^{i_1}x.
	\end{align*}
	For each $k\in[n]$, defining
	\begin{align*}
	y_{n,k}(r) = \frac{1}{n^k}\sum_{i \in \Delta_{\text{disc}}^k(n)} \Big(\Pi_{m=k+1}^2M^{i_m}\cL_{n,r}\Big)M^{i_1}x,
	\end{align*}
	we see by Lemma~\ref{lem:ZenoWorkhorse} that
	\begin{align*}
	\lim_{n\to\infty} y_{n,k}(r) = \frac{\left(P\cL_r P\right)^k}{k!}x
	\end{align*}
 uniformly in $r\in R'.$
	Moreover, by using the facts that $\|M_r\|\le1$ and $\sup_{n\in\N,r\in R}\|\cL_{n,r}\|\le C$ for some finite $C>0$, together with the fact that
 \begin{align*}
     \frac{|\Delta^k_{\text{disc}}(n)|}{n^k}\le \frac{1}{k!},
 \end{align*}
 proven in \eqref{eq:SimplUniUpperBound}, we see that
	\begin{align*}
	\|y_{n,k}(r)\| \le \frac{C^k}{k!} \|x\|
	\end{align*}
	Hence, by the dominated convergence theorem we obtain
	\begin{align*}
	\lim_{n\to\infty}\left(M_r+\frac{\cL_{n,r}}{n}\right)^nx &=\lim_{n\to\infty}\Big( M_r^nx + \sum_{k=1}^n y_{n,k} (r)\Big)=Px + \sum_{k=1}^\infty \frac{\left(P\cL_r P\right)^k}{k!}x = e^{P\cL_r P}Px,
	\end{align*}
 with convergence again uniformly in $r\in R'.$
 To prove the quantitative bound on the speed of convergence \eqref{eq:SpeedBino}, we use \eqref{eq:limitpimitBino} in Lemma~\ref{lem:ZenoWorkhorse} to note that for every $\mathbf{N} =(N_l)_{l\in\N}\subseteq\N$ with $N_1\le n$ we have
  \begin{equation}
    \begin{split}
		&\left\|y_{n,k}(r)  -\frac{(P\cL_r P)^k}{k!}x\right\| \le\frac{kC^k}{(k-1)!}\Big(\,s'_n\|x\| + s^{\max}_n\left(x,\cL_r,\mathbf{N}_{k+1}\right)\Big),
  \end{split}
	\end{equation} 
   for some $C\ge1,$ where $\mathbf{N}_{k+1} =(N_1,\dots,N_{k+1}),$
and \begin{equation}
s^{\max}_n(x,\cL_r,\mathbf{N}_{k+1})=\max_{l\in[k+1]}\left(\left(\sup_{\tilde r\in R}\|\cL_{\tilde r}\|\right)^{-l+1}s_{N_l}\left((\cL_rP)^{l-1}x\right) +\frac{N_l}{n}\|x\|\right).
 \end{equation}
Using again $\|\cL_r\| \le C$ this gives
 \begin{align}
 \label{eq:FinalBinoBound}
\nn&\left\|\left(\left(M_r+\frac{\cL_{n,r}}{n}\right)^n- e^{P\cL_r P}P\right)x\right\| \\&\le\nn \left\|\left(M^n_r-P\right)x\right\| +\sum_{k=1}^n\left\|y_{n,k}(r) - \frac{\left(P\cL_r P\right)^k}{k!}\right\| + \sum_{k=n+1}^\infty\frac{C^k}{k!}\|x\| \\&\nn\le s_n(x) + \sum_{k=1}^n\frac{kC^k}{(k-1)!}\Big(\,s'_n\|x\| + s^{\max}_n\left(x,\cL_r,\mathbf{N}_{k+1}\right)\Big) + \frac{Ce^C}{n}\|x\| \\&\le4C^2e^C \Big(s'_n\|x\| + s^{\sup}_n(x,\cL_r,\mathbf{N})\Big).
 \end{align}
 where we have used $n\mapsto s_n(x)$ being monotonically decreasing and $N_1\le n$ and
 \comment{
 \begin{align}
\left\|\sum_{k=0}^n\frac{\left(P\cL_rP\right)^k}{k!}-e^{P\cL_r P}\right\| \le  \sum_{k=n+1}^\infty\frac{C^k}{k!} \le \frac{C}{n+1}\sum_{k=n}^\infty\frac{C^k}{k!} \le\frac{Ce^C}{n}
 \end{align}}
 \begin{align*}
     \sum_{k=1}^\infty\frac{kC^k}{(k-1)!} =C\frac{d}{dx}\Big|_{x=C}\left(x \sum_{k=0}^\infty  \frac{x^k}{k!}\right) \le 2C^2e^C.
 \end{align*}
 and have defined
 \begin{equation}
s^{\sup}_n(x,\cL_r,\mathbf{N})=\sup_{l\in\N}\left(\left(\sup_{\tilde r\in R}\|\cL_{\tilde r}\|\right)^{-l+1}s_{N_l}\left((\cL_rP)^{l-1}x\right) +\frac{N_l}{n}\|x\|\right).
 \end{equation}
 Renaming again the constant $C$ in \eqref{eq:FinalBinoBound} gives the desired bound.

\end{proof}

\section{Summary and open problems}
\label{sec:Summary}

We proved quantum Zeno limits and strong damping limits in strong topology in a unified way (cf.~Theorems~\ref{thm:StrongZenoGeneralQuant} and~\ref{thm:StrongDamping}) and while merely demanding that the mixing limits $M^n\xrightarrow[n\to\infty]{} P$ and $e^{\gamma\cK}\xrightarrow[\gamma\to\infty]{}P$ hold in strong topology. Both results are quantitative and contain a bound on the speed of convergence in terms of the convergence speed of the respective mixing limits. This is the first time that a bound on the speed of convergence for the quantum Zeno limit has been obtained in this setup. Furthermore, our result on the strong damping limit can be seen as the first for unbounded generators $\cK.$

For the unified proof we introduced a novel operator product, which we called the generalised binomial product, which generalises both the Zeno product and damped evolution operator. We then analysed its asymptotic behaviour in Theorem~\ref{thm:StrongBinoFormula}.

Using these results, we were able to prove quantum Zeno- and strong damping limits for the attenutator channel and semigroup with a bound on the corresponding speeds of convergence (Example~\ref{ex:Atten} and Proposition~\ref{prop:AttZenoDamp}).

\bigskip

An important open problem which could be the content of future research is the following:
\comment{Another important open problem is to extend the results of Zeno- and strong damping limits for unbounded generators $\cL.$}
    The articles \cite{BeckerDattaSalz_Zeno_2021,möbus2022optimal,Tim_TrotterZeno_2024} generalised, under the spectral condition on $M$, the results on convergence of the Zeno product $\left(Me^{t\cL/n}\right)^n$ partially to unbounded generators $\cL,$ though only under %however under 
    fairly strict constraints. It would be interesting to also provide extensions of our results on Zeno- and strong damping limits in strong topology, Theorem~\ref{thm:StrongZenoGeneralQuant} and~\ref{thm:StrongDamping}, for unbounded $\cL.$ This could potentially be achieved under some form of energy constraint on the input state, reminiscent of the assumption of finite particle number in Proposition~\ref{prop:AttZenoDamp}. In the case of strong damping, it would be particularly desirable to obtain a rigorous analysis of the proposed paradigm for photonic dynamically error corrected qubits from \cite{Mirrahimi_DynamicErrorCatQubits_2014,Mirri_New_2019,Mirrahimi_LectureNotes_2023}.

\bigskip 

\noindent\textbf{Acknowledgments.} The author would like to thank Simon Becker, Daniel Burgarth, Nilanjana Datta, Hamza Fawzi, Omar Fawzi, Niklas Galke, Paul Gondolf, Tim Möbus, Cambyse Rouz\'e and Michael Wolf for helpful discussions. The author acknowledges funding from the Cambridge Commonwealth,
European and International Trust, from the European Research Council (ERC Grant AlgoQIP, Agreement No. 851716) and from a government grant managed by the Agence Nationale de la Recherche under the Plan France 2030 with the reference ANR-22-PETQ-0006.

\bibliography{Ref}
\bibliographystyle{alpha}

\end{document}